\documentclass[%
 reprint,
superscriptaddress,
nofootinbib,
 amsmath,amssymb,
 aps,
]{revtex4-2}

\usepackage[dvipsnames]{xcolor}
\usepackage{graphicx}% Include figure files
\usepackage{dcolumn}% Align table columns on decimal point
\usepackage{upgreek}
\usepackage{xspace}

\usepackage{bm}% bold math
\begin{document}
% $Id: commands.tex 934 2013-06-19 20:56:45Z mfloris $

\newcommand{\ITS}          {\mathrm{ITS}}
\newcommand{\TOF}          {\mathrm{TOF}}
\newcommand{\ZDC}          {\mathrm{ZDC}}
\newcommand{\ZDCs}         {\mathrm{ZDCs}}
\newcommand{\ZNA}          {\mathrm{ZNA}}
\newcommand{\ZNC}          {\mathrm{ZNC}}
\newcommand{\SPD}          {\mathrm{SPD}}
\newcommand{\SDD}          {\mathrm{SDD}}
\newcommand{\SSD}          {\mathrm{SSD}}
\newcommand{\TPC}          {\mathrm{TPC}}
\newcommand{\VZERO}        {\mathrm{V0}}
\newcommand{\VZEROA}       {\mathrm{V0-A}}
\newcommand{\VZEROC}       {\mathrm{V0-C}}
\newcommand{\pip}          {$\pi^{+}$}
\newcommand{\pim}          {$\pi^{-}$}
\newcommand{\kap}          {K$^{+}$}
\newcommand{\kam}          {K$^{-}$}
\newcommand{\he}{$^{3}{\mathrm{He}}$}
\newcommand{\ahe}{$^{3}\overline{\rm He}$}
\newcommand{\heFour}{$^{4}{\mathrm{He}}$}
\newcommand{\aheFour}{$^{4}\overline{\rm He}$}

\newcommand{\ad}{$\overline{\rm d}$}
\newcommand{\pbar}         {\ensuremath{\overline{\mathrm{p}}}}
\newcommand{\nbar}         {\ensuremath{\overline{\mathrm{n}}}}
\newcommand{\kzero}        {\ensuremath{\mathrm{ K^{0}_{S}}}}
\newcommand{\vzero}        {\ensuremath{\mathrm{V}^0}}
\newcommand{\lmb}          {\ensuremath{\Lambda}}
\newcommand{\almb}         {\ensuremath{\bar{\Lambda}}}
\newcommand{\allpart}      {$\pi^{\pm}$, K$^{\pm}$, \kzero, p(\pbar) and \lmb(\almb)}
\newcommand{\allpikp}      {$\pi^{\pm}$, K$^{\pm}$ and p(\pbar)}
\newcommand{\pikp}         {$\pi$, K, and p}
\newcommand{\allpi}        {$\pi^{\pm}$}
\newcommand{\allk}         {K$^{\pm}$}
\newcommand{\allp}         {p(\pbar)}
\newcommand{\alllmb}       {\lmb(\almb)}
\newcommand{\degree}       {$^{\mathrm{o}}$}
\newcommand{\dg}           {\mbox{$^\circ$}}
\newcommand{\dedx}         {\ensuremath{\mathrm{d}E/\mathrm{d}x}}
\newcommand{\dndy}         {d$N$/d$y$}
\newcommand{\pp}           {pp}
\newcommand{\ppbar}        {\mbox{$\mathrm {p\overline{p}}$}}
\newcommand{\PbPb}         {\mbox{Pb--Pb}}
\newcommand{\pPb}          {\mbox{p--Pb}}
\newcommand{\AuAu}         {\mbox{Au--Au}}
\newcommand{\pseudorap}    {\mbox{$\left | \eta \right | $}}
\newcommand{\dNdeta}       {\ensuremath{\mathrm{d}N_\mathrm{ch}/\mathrm{d}\eta}}
\newcommand{\dNdy}         {\ensuremath{\mathrm{d}N_\mathrm{ch}/\mathrm{d}y}}
\newcommand{\dNdyst}       {\ensuremath{\sqrt{\frac{dN_\pi/dy}{s_T}}}}
\newcommand{\dNdetatr}     {\mathrm{d}N_\mathrm{tracklets}/\mathrm{d}\eta}
\newcommand{\dNdetar}[1]   {\mathrm{d}N_\mathrm{ch}/\mathrm{d}\eta\left.\right|_{|\eta|<#1}}
\newcommand{\dNdetamean}   {\ensuremath{\langle\dNdeta  \rangle}}
\newcommand{\lum}          {\, \mbox{${\mathrm{ cm}}^{-2} {\mathrm {s}}^{-1}$}}
\newcommand{\barn}         {\, \mbox{${\mathrm{barn}}$}}
\newcommand{\m}            {\, \mbox{${\mathrm{m}}$}}
\newcommand{\ncls}         {\ensuremath{N_{cls}}}
\newcommand{\nsigma}       {\ensuremath{n\sigma}}
\newcommand{\dcaxy}        {\ensuremath{\mathrm{DCA_{\rm xy}}}}
\newcommand{\dcaz}         {\ensuremath{\mathrm{DCA_{\rm z}}}}
\newcommand{\EcrossB}      {E$\times$B}%{\ensuremath{{\rm E}\times{\rm B}}}
\newcommand{\bb}           {Bethe-Bloch}
\newcommand{\s}            {\ensuremath{\sqrt{s}}}
\newcommand{\PT}           {\ensuremath{p_{\mathrm{T}}}}
\newcommand{\MT}           {\ensuremath{m_{\mathrm{T}}}}
\newcommand{\hlab}         {\ensuremath{\eta_{\mathrm{lab}}}}
\newcommand{\ynn}         {\ensuremath{y_{\mathrm{NN}}}}
\newcommand{\ycms}         {\ensuremath{y_{\mathrm{CMS}}}}
\newcommand{\ylab}         {\ensuremath{y_{\mathrm{lab}}}}
\newcommand{\ppi}          {\ensuremath{{\mathrm{p}}/\pi}}
\newcommand{\kpi}          {\ensuremath{{\mathrm{K}}/\pi}}
\newcommand{\lpi}          {\ensuremath{{\mathrm{ \Lambda}}/\pi}}
\newcommand{\lks}          {\ensuremath{{\mathrm{ \Lambda}}/\mathrm{{K}^{0}_{S}}}}
\newcommand{\mt}           {\ensuremath{m_{\mathrm{T}}}}
\newcommand{\snn}          {\ensuremath{\sqrt{s_{\mathrm{NN}}}}}
\newcommand{\snnbf}        {\ensuremath{\mathbf{{\sqrt{s_{\mathbf{ NN}}}}}}}
\newcommand{\sonly}        {\ensuremath{\sqrt{s}}}
\newcommand{\Npart}        {\ensuremath{N_\mathrm{part}}}
\newcommand{\avNpart}      {\ensuremath{\langle N_\mathrm{part} \rangle}}
\newcommand{\avNpartdata}  {\ensuremath{\langle N_\mathrm{part}^{\mathrm{data}} \rangle}}
\newcommand{\Ncoll}        {\ensuremath{N_\mathrm{coll}}}
\newcommand{\avNcoll}      {\ensuremath{\langle N_\mathrm{coll} \rangle}}
\newcommand{\Dnpart}       {\ensuremath{D\left(\Npart\right)}}
\newcommand{\DnpartExp}    {\ensuremath{D_{\mathrm{exp}}\left(\Npart\right)}}
\newcommand{\dNdetapt}     {\ensuremath{\dNdeta\,/\left(0.5\Npart\right)}}
\newcommand{\dNdetaptr}[1] {\ensuremath{\dNdetar{#1}\,/\left(0.5\Npart\right)}}
\newcommand{\dNdetape}     {\left(\ensuremath{\dNdeta\right)/\left(\avNpart/2\right)}}
\newcommand{\dNdetaper}[1] {\ensuremath{\dNdetar{#1}\,/\left(\avNpart/2\right)}}
\newcommand{\dndydpt}      {\ensuremath{{\mathrm{d}}^2N/({\mathrm{d}}y {\mathrm{d}}p_{\mathrm{t}})}}
\newcommand{\abs}[1]       {\ensuremath{\left|#1\right|}}
\newcommand{\signn}        {\ensuremath{\sigma^{\mathrm{inel.}}_{\mathrm{NN}}}}
\newcommand{\vz}           {\ensuremath{V_{z}}}
\newcommand{\Tfo}          {\ensuremath{{T}_{\mathrm{kin}}}}
\newcommand{\Tch}          {\ensuremath{{T}_{\mathrm{ch}}}}
\newcommand{\bT}           {\ensuremath{\beta_{\mathrm{T}}}}
\newcommand{\avbT}         {\ensuremath{\langle \beta_{\mathrm{T}}\rangle}}
\newcommand{\avpT}         {\ensuremath{\langle \PT \rangle}}
\newcommand{\muB}          {\ensuremath{\mu_{B}}}
\newcommand{\stat}         {({\it stat.})}
\newcommand{\syst}         {({\it sys.})}
\newcommand{\Fig}[1]       {Fig.~\ref{#1}}
\newcommand{\green}[1]     {\textcolor{green}{#1}}
\newcommand{\blue}[1]      {\textcolor{blue}{#1}}
\newcommand{\red}[1]       {\textcolor{red}{#1}}
\newcommand{\white}[1]     {\textcolor{white}{#1}}
\newcommand{\gevc}         {\ensuremath{{\mathrm{GeV}}/c}}
\newcommand{\gevcsq}       {\ensuremath{{\mathrm{GeV}^4}/c^2}}
\newcommand{\mevc}         {\ensuremath{{\mathrm{MeV}}/c}}
\newcommand{\avg}[1]       {\ensuremath{\left\langle#1\right\rangle}}
\newcommand{\tkin}         {\ensuremath{T\mathrm{_{kin}}}}
\newcommand{\rppb}         {\ensuremath{R\mathrm{_{pPb}}}}
\newcommand{\rdau}         {\ensuremath{R\mathrm{_{dAu}}}}
\newcommand{\mtof}         {\ensuremath{{m_{\mathrm{TOF}}^2}}}
\newcommand{\dbard}        {\ensuremath{\mathrm{\bar{d}/d}}}
\newcommand{\pbarp}        {\ensuremath{\mathrm{\bar{p}/p}}}
\newcommand{\nbarn}        {\ensuremath{\mathrm{\bar{n}/n}}}
\newcommand{\alphas}       {\ensuremath{\alpha_{\mathrm{S}}}}
\newcommand{\mub}          {\ensuremath{\mathrm{\mu}_{\mathrm{B}}}}
\newcommand{\Btwo}         {\ensuremath{B_2}}
\newcommand{\geant}  {${\textsc{Geant4}}$\xspace}
\newcommand{\sigmainel}    {\ensuremath{\sigma_{\mathrm{inel}}^{\bar{\rm d}}}}

%\renewcommand{\labelitemi} {$-$}
%==========================================================%
%%% inline warnings for internal discussion 
%\newcommand{\warn}[1]      {\textbf{\textcolor{red}{[#1]}}}
\newcommand{\warn}[1]      {{\small\textbf{\textcolor{red}{(!\footnote{\textbf{(!)}~#1})}}}}
\newcommand{\warnin}[1]         {\textit{\textcolor{red}{(#1)}}}
%\newcommand{\warn}[1]      {#1}
%\newcommand{\warn}[1]      {{\small\textbf{(!\footnote{\textbf{(!)}~#1})}}\marginpar{\textbf{---}}}
%\newcommand{\todo}[1]      {\textbf{\textcolor{red}{[TODO: #1]}}}
%%% fake numbers
\newcommand{\fake}[1]      {\textbf{\textcolor{red}{#1}}}
\newcommand{\final}[1]     {\textbf{\textcolor{blue}{#1}}}
\newcommand{\prelim}[1]    {\textbf{\textcolor{magenta}{#1}}}
\renewcommand{\mod}[1]       {\textbf{\textcolor{red}{#1}}}

% to count the words
\newcommand{\detailtexcount}[1]{%
  \immediate\write18{texcount -merge -sum -q #1.tex output.bbl > #1.wcdetail }%
  \verbatiminput{#1.wcdetail}%
  \newpage
}

%Johannes:
\newcommand{\dbar}{\ensuremath{{\bar d}}}
\newcommand{\eV}[0]{\,\mathrm{eV}}
\newcommand{\keV}[0]{\,\mathrm{keV}}
\newcommand{\MeV}[0]{\,\mathrm{MeV}}
\newcommand{\GeV}[0]{\,\mathrm{GeV}}
\newcommand{\TeV}[0]{\,\mathrm{TeV}}
\newcommand{\sigmav}[0]{\ensuremath{\langle \sigma v \rangle}}
\newcommand{\kpc}{\,\mathrm{kpc}}

\newcommand{\jc}[1]{{\color{blue}[JH: #1]}}
\newcommand{\lf}[1]{{\color{magenta}[LF: #1]}}
\newcommand{\ls}[1]{{\color{red}[LS: #1]}}
\newcommand{\sk}[1]{{\color{red}[SK: #1]}}
\newcommand{\tp}[1]{{\color{green}[#1]}}
%need to decide which convention to use for \pbar. I like mine better, obvs. 

%thomas
\newcommand{\addref}[0]{{\color{red}$\left[\mathrm{Add\,reference}\right]$}}

% annoying abbreviations used in ADS references
\def\amp{\&}
\def\nat{Nature}
\def\pra{Phys. Rev. A}
\def\prb{Phys. Rev. B}
\def\prc{Phys. Rev. B}
\def\prd{Phys. Rev. D}
\def\mnras{MNRAS}
\def\apj{ApJ}
\def\apjl{ApJL}
\def\apjs{ApJS}
\def\aap{A\&A}
\def\aaps{A\&A Supp.}
\def\aapr{A\&A Rev.}
\def\araa{ARA\&A}
\def\aj{AJ}
\def\qjras{QJRAS}
\def\physrep{Phys. Rep.}
\def\apss{Ap\&SS}               % Astrophysics and Space Science
\def\planss{P\&SS}              % Planetary and Space Science
\def\pasp{Publ. Astr. Soc. Pac.}
\def\jcp{J. Chem. Phys.}
\def\pasj{Pub. Astron. Soc. Japan}
\def\ssr{Space Science Reviews}
\def\grl{Geophysical Research Letters}
\def\jgr{Journal of Geophysical Research}
\def\nar{New Astronomy Reviews}
\def\jcap{JCAP}
\def\sovast{Sov. Astron.}
\preprint{APS/123-QED}

\title{Reevaluation of the cosmic antideuteron flux from\\  cosmic-ray interactions and from exotic sources}% Force line breaks with \\

\author{Laura Šerkšnytė}
 \email{laura.serksnyte@tum.de}
\affiliation{Department of Physics, Technical University of Munich, James-Franck-Straße, 85748 Garching, Germany} %
\author{Stephan K\"onigstorfer}%
 \email{s.koenigstorfer@tum.de}
\affiliation{Department of Physics, Technical University of Munich, James-Franck-Straße, 85748 Garching, Germany} %
\author{Philip von Doetinchem}%
\affiliation{Department of Physics and Astronomy, University of Hawaii at Manoa,
2505 Correa Road, Honolulu, Hawaii 96822, USA} %
\author{Laura Fabbietti}%
\affiliation{Department of Physics, Technical University of Munich, James-Franck-Straße, 85748 Garching, Germany} %
\author{Diego Mauricio Gomez-Coral}%
\affiliation{Department of Physics and Astronomy, University of Hawaii at Manoa,
2505 Correa Road, Honolulu, Hawaii 96822, USA} %
\author{Johannes Herms}%
\affiliation{Max-Planck-Institut f\"ur Kernphysik, Saupfercheckweg 1, 69117 Heidelberg, Germany} %
\author{Alejandro Ibarra}%
\affiliation{Department of Physics, Technical University of Munich, James-Franck-Straße, 85748 Garching, Germany} %
\author{Thomas P\"oschl}%
\affiliation{Department of Physics, Technical University of Munich, James-Franck-Straße, 85748 Garching, Germany} %
\author{Anirvan Shukla}%
\affiliation{Department of Physics and Astronomy, University of Hawaii at Manoa,
2505 Correa Road, Honolulu, Hawaii 96822, USA} %
\author{Andrew Strong}%
\affiliation{Max-Planck-Institut f\"ur extraterrestrische Physik, Giessenbachstraße,
85748 Garching, Germany} %
\author{Ivan Vorobyev}%
\affiliation{Department of Physics, Technical University of Munich, James-Franck-Straße, 85748 Garching, Germany} %

\date{\today}% It is always \today, today,

\begin{abstract}
Cosmic-ray antideuterons could be a key for the discovery of exotic phenomena in our Galaxy, such as dark-matter annihilations or primordial black hole evaporation. Unfortunately the theoretical predictions of the antideuteron flux at Earth are plagued with uncertainties from the mechanism of antideuteron production and propagation in the Galaxy. We present the most up-to-date calculation of the antideuteron fluxes from cosmic-ray collisions with the interstellar medium and from exotic processes. We include for the first time the antideuteron inelastic interaction cross section recently measured by the ALICE collaboration to account for the loss of antideuterons during propagation. In order to bracket the uncertainty in the expected fluxes, we consider several state-of-the-art models of antideuteron production and of cosmic-ray propagation.
\end{abstract}

\maketitle

%\tableofcontents
\section{Introduction}

In recent years there has been a growing interest in searches for cosmic-ray antiparticles with space-based and balloon-borne experiments, like BESS, PAMELA, and AMS-02 \cite{Fuke:2005it,2012PhRvL.108m1301A,besspbar, Adriani:2012paa, PAMELA:2017bna,1998ApJ...509..212S, AMS:2016oqu}. 
One of the motivations is that rare antiparticles act as messengers for exotic processes in the Galaxy, such as dark-matter annihilation or decay \cite{Donato:1999gy,Baer:2005tw,Donato:2008yx,Duperray:2005si,Ibarra:2012cc,Ibarra2013a,Fornengo:2013osa,Dal:2014nda, Korsmeier:2017xzj,Tomassetti:2017qjk,Lin:2018avl,Li:2018dxj,Cirelli:2014qia, Carlson:2014ssa, Coogan:2017pwt}, which have a very low astrophysical background at kinetic energies below few GeV/nucleon. This background is generated by interactions of primary cosmic rays, like protons or $\upalpha$-particles, with the interstellar medium (ISM). The only cosmic-ray antiparticles that have been detected to date are positrons \cite{pamela,ams02first, ams4} and antiprotons \cite{Fuke:2005it,besspbar,2012PhRvL.108m1301A,Adriani:2012paa, AMS:2016oqu}. The exquisite measurements of the positron and antiproton spectra are  currently being actively  interpreted and analyzed. 
The positron energy spectrum shows a hardening at high energies that cannot be explained by standard cosmic-ray propagation models, like \textsc{galprop}~\cite{1998ApJ...493..694M}. Explanations span a wide range of very different models from various acceleration mechanisms to positron production in pulsars and dark-matter annihilation~\cite{2015A&A...575A..67B}. For antiprotons, deviations from the standard cosmic-ray propagation prediction have also been found  with varying degrees of significance~\cite{PhysRevLett.118.191101,PhysRevLett.118.191102,Reinert:2017aga,Cuoco:2019kuu,Cholis:2019ejx,Boudaud:2019efq, Heisig:2020nse}. Some of these antiproton studies suggest a dark-matter interpretation that can also explain the $\upgamma$-ray excess observed in the Galactic Center~\cite{PhysRevLett.118.191102,Cholis:2019ejx}. However the deviations of the observed antiproton fluxes from the non-exotic background predictions are not very large. This poses a challenge for the interpretations, influenced by other uncertainties related to, e.g., antiparticle production cross sections for primary cosmic-ray interactions with the ISM, propagation parameters, solar modulation, or instrumental resolution effects. \\
In the search for cosmic messengers with a higher signal from exotic processes with respect to the astrophysical background, cosmic-ray antideuterons were proposed as a possible signature about 20\,years ago~\cite{Donato:1999gy}, for recent reviews see~\cite{Aramaki:2015pii,review2020}. Antideuterons are composed of one antiproton and one antineutron. From studies of light antinuclei production at particle colliders on Earth, it is known that the addition of one antinucleon to the antinucleus suppresses the production cross section by about a factor of 1000 in proton--proton collisions~\cite{ALICE:2017xrp}, resulting in the prediction of a much lower cosmic antideuteron flux with respect to antiprotons. Thus, in a fairly generic way, it is possible in a broad range of models, like dark-matter annihilation, to produce a cosmic antideuteron flux that is both several orders of magnitude above the astrophysical antideuteron background prediction and shows a different spectral shape with respect to the expected antideuteron background due to the underlying kinematics, but is also several orders of magnitude lower with respect to antiprotons. 
The detection of only one or a few cosmic antideuterons would be a potential breakthrough  for finding imprints of exotic processes because they can be easily separated from other background antideuterons. The search requires experiments with large acceptance, long measurement time, and high particle identification power. The best exclusion limits have been reported by the BESS experiment collaboration~\cite{Fuke:2005it}, and the analysis of the currently operational multi-purpose AMS-02 experiment aboard the International Space Station is ongoing. The upcoming balloon-borne experiment GAPS is a dedicated low-energy cosmic antinuclei search and is expected to have its first flight by the end of 2022~\cite{Aramaki:2014oda,Aramaki:2015laa,GAPS:2020axg}.
Interestingly, the AMS-02 collaboration reported several antihelium candidate events~\cite{Kounine:2019gsh}, which already generated broad theoretical interest, with no preferred explanation. However, the consensus seems to be that if confirmed, it would have a transformative impact on understanding the processes in the Galaxy.  A confirmed detection of antihelium would also directly impact the predictions of the antideuteron flux. Therefore, any model explaining the potential cosmic antihelium signal must not be above the measured antiproton flux and must respect current antideuteron detection limits.\\
The properties of light cosmic-ray antinuclei have been studied in accelerator-based experiments using various colliding systems and colliding energies to determine their production mechanisms quantitatively~\cite{SimonGillo:1995dh,Armstrong:2000gd,Afanasev:2000ku,Anticic:2004yj,Adler:2004uy,Alper:1973my,Henning:1977mt,Alexopoulos:2000jk,Aktas:2004pq,Asner:2006pw,ALEPH:2006qoi,ALICE:2015wav,Adam:2015yta,Acharya:2017dmc,ALICE:2019dgz,Agakishiev:2011ib,Abelev:2010,Adam:2015pna,Adam:2019phl}. Despite of the plethora of very precise measurements, the studies do not suffice to constrain the antinuclei production cross sections within the very wide energy range of collisions occurring between high-energy cosmic rays and nuclei present in the interstellar medium. Hence relatively large uncertainties of about a factor of 10 are still present in the most critical energy region~\cite{Gomez-Coral:2018yuk}. 
 Thus more measurements of this kind are needed, and especially more comprehensive modelling of the antinuclei production in hadron--hadron collisions is necessary~\cite{Blum:2019suo,Kachelriess:2020uoh}. 
Accelerator-based experiments  not only constrain the yield of secondary antinuclei in our Galaxy but also set essential boundaries on the dark-matter annihilation processes resulting in the production of antinuclei (eg.~\cite{Ibarra:2012cc,Winkler:2021cmt}).
Another crucial aspect that can be studied at accelerators is the inelastic interaction cross section of light antinuclei with matter. Such processes play a fundamental role in the propagation of antinuclei for cosmic-ray experiments. Recently measurements of the inelastic cross section of antideuterons \cite{ALICE:2020zhb} have been performed over a wide momentum range, and they can now be used as input to propagation programs. 
 Several additional experimental efforts are underway to reduce the related uncertainties using new data from the ALICE~\cite{ALICE:2020zhb} and NA61/SHINE~\cite{na61} experiments. \\
  After introducing potential antinuclei sources (Sec.~\ref{s-sources}), this study focuses on using the latest data and models for antinuclei production and interaction cross sections together with up-to-date cosmic-ray propagation models (Secs.~\ref{s-prop}, \ref{s-cross}) for a prediction of exotic signal and background antideuteron fluxes.
\section{Cosmic-Ray Propagation}
\label{s-prop}
Relativistic nuclei, and electrons and positrons, pervade our Galaxy and are collectively known as cosmic rays. They span energies from MeV to PeV and above.
 Detailed expositions of cosmic-ray transport is given in \cite{2007ARNPS..57..285S,GrenierBlackStrong2015}; here we summarize the essential concepts.
Galactic cosmic rays are thought to originate mainly from diffusive shock acceleration of interstellar gas by supernova remnants, with other sources like pulsars and pulsar wind nebulae possibly also contributing.%~\cite{Laura-Andrew}.

Cosmic rays from such sources are referred to as primary; they interact with the hydrogen (H) and helium (He) atoms within the interstellar medium to produce secondary nuclei and antinuclei. For example, a  spallation process converts a primary carbon nucleus (C, $Z=6$, $A=12,13$) into a secondary boron nucleus (B, $Z=5$, $A=10,11$). Cosmic-ray nuclei cover the full range of isotopes from H through He to Ni, and secondary antinuclei and antideuterons. 

Cosmic rays propagate in the Galaxy mainly through diffusion due to scattering on interstellar turbulence and convection by Galactic winds. They eventually leave the Galaxy, filling a region known as the cosmic-ray halo which has a vertical extent of several\,kpc. Their residence time in the Galaxy is about 10-100 Myr. Their energy is lost by hadronic interactions such as pion production and ionization.  Inelastic collisions and radioactive decay are further loss mechanisms. The cosmic-ray propagation is described by the Fokker-Planck equation which can be written as
\begin{widetext}
\begin{equation}
\label{eqn:TransportEquation}
\frac{\partial\psi}{\partial t} =Q(\textbf{\textit{r}},p) + \textbf{div}(D_{\mathrm{xx}}\textbf{grad}\psi-\textbf{\textit{V}}\psi)+\frac{\partial}{\partial p}p^2D_{\mathrm{pp}}\frac{\partial }{\partial p}\frac{\psi}{p^2}-\frac{\partial}{\partial p} \left[ \psi \frac{\mathrm{d}p}{\mathrm{d}t}-\frac{p}{3}(\textbf{div}\cdot \textbf{\textit{V}})\psi\right]-\frac{\psi}{\tau},
\end{equation}
\end{widetext}
where $\psi=\psi(\textbf{\textit{r}}, p, t)$ is the time-dependent cosmic-ray density per unit of the total particle momentum at position $\textbf{\textit{r}}$. $Q(\textbf{\textit{r}},p)$ is the source term of the cosmic rays, which can include primary particles injected by  supernova remnants, secondaries coming from spallation and cosmic-ray collisions with the interstellar medium as well as more exotic sources such as dark-matter annihilation. $D_{\mathrm{xx}}$, $\textbf{V}$, and $D_{\mathrm{pp}}$ are the spatial diffusion coefficient, the convection velocity, and the diffusive re-acceleration coefficient, respectively, and are called propagation parameters. The last term $\psi$/$\tau$ accounts for particles lost via decay, fragmentation and inelastic interactions in the Galaxy.

Solar modulation is significant below about 1 GeV/nucleon kinetic energy per nucleon and is treated separately using the Force Field approximation~\cite{Gleeson:1968zza} or with special codes such as HelMod~\cite{BOSCHINI20182859,BOSCHINI20192459}.

In this work the \textsc{galprop} program \cite{1998ApJ...509..212S,2020ApJS..250...27B,J_hannesson_2021} is used to compute cosmic-ray propagation. In this setup, our Galaxy is approximated as a cylinder with a halo height of 4\,kpc and a radius of 20\,kpc.
The distribution of cosmic-ray sources in the Galaxy is based on supernova remnants, but pulsars are used since the supernova distribution is very uncertain. Pulsars should have a similar distribution to supernova remnants, but more observations of pulsars are available, and their distances are measured more precisely. The source distribution is thus parameterized as a function of galactocentric radius based on pulsars and an exponential behaviour above and below the Galactic plane with a scale length of order 100 pc. The cosmic-ray injection spectrum can be parameterized as a power-law in momentum, $p^{-\alpha}$, where the spectral index $\alpha$ is momentum dependent and assumes typical values of $\alpha=2.3$ above a few GeV, flattening to 1.8 at lower momenta. Isotopic abundances of primary cosmic rays are fixed in \textsc{galprop} by the cosmic-ray flux values measured at 100 GeV/nucleon kinetic energy because solar modulation does not affect this energy range. 
The halo height $z_{\rm h}$, at which cosmic rays are assumed to fall to zero, is a free parameter with values between 1 and 20\,kpc, and it is determined from the fit to cosmic-ray data carried out to fix the propagation parameters.
The propagation parameters are determined by exploiting the relation between primary and secondary cosmic rays observed by experiments on balloons and satellites. Since the primary cosmic-ray spectrum can be measured and the cross sections for the secondary production are known, the primary-to-secondary ratios can be used to constrain the propagation parameters.
The parameters are determined by fitting the primary H and He spectra, the boron-to-carbon ratio (B/C), and other secondary-to-primary ratios~\cite{1998ApJ...509..212S}.

Convection by the Galactic wind is specified by the convection velocity, which is assumed to have a linear increase with distance from the plane. It must be equal to 0 at the Galactic plane to avoid a discontinuity there. The diffusion coefficient can be parameterized as $D_{\rm xx}=\beta R^{-\alpha}$ where $\beta=v/c$, $v$ is velocity, $c$ is the speed of light and $R=pc/eZ$ is the rigidity. The $\beta$ term reflects that diffusion depends on the speed with which particles scatter on interstellar turbulence.
Astronomical information such as the distribution of atomic and molecular gas in the Galaxy for spallation, the interstellar radiation fields for leptonic interactions are based on large-scale surveys at various levels of detail.

The present work is based on \textsc{galprop} v56~\cite{2020ApJS..250...27B} with various modifications because the propagation of antinuclei (apart from antiprotons) was not included in \textsc{galprop} before this work. Full technical details of the \textsc{galprop} parameters and solution methods, with diagnostics of the solution, together with illustrative analytical solutions, can be found in the \textsc{galprop} Explanatory Supplement\footnote{https://gitlab.mpcdf.mpg.de/aws/galprop}.

Despite the complex framework of \textsc{galprop},
it is important to note that secondary production of antiprotons and antideuterons is determined mainly by the (momentum-dependent) "grammage" or column of matter traversed, not directly by the halo size, since the latter only affects radioactive species via the residence time. The well-measured B/C ratio as a function of energy determines the grammage. Any combination of $D_{\rm{xx}}$ and halo size, which gives the correct B/C can be used, at least to a good approximation; B/C constrains only\footnote{https://gitlab.mpcdf.mpg.de/aws/galprop: Explanatory Supplement, equation (63)} $z_{\rm h}/D_{\rm{xx}}$.

In this work, the relevant antideuteron source functions and the characteristic inelastic cross sections have been implemented in \textsc{galprop} to estimate the antideuteron fluxes. The details can be found in sections~\ref{s-sources} and~\ref{s-cross}. The propagation parameters obtained in Boschini et al.(P-scenario)~\cite{2020ApJS..250...27B} (Table~\ref{tab:tablePar}) have been used, and the kinetic-energy grid employed in \textsc{galprop} has been adapted to the available antideuteron production cross sections. To illustrate the uncertainty of \textsc{galprop} propagation parameters, the Cuoco et al.~parameterization is employed as well (Table~\ref{tab:tablePar}).

\begin{table}[b]
\caption{\label{tab:tablePar}%
The main propagation parameters used in \textsc{galprop}. For full parameter sets refer to Boschini et al. and Cuoco et al.
}
\begin{ruledtabular}
\begin{tabular}{lccc}
 Parameter & Units &Boschini et al. & Cuoco et al.\\ [0.5ex] 
\colrule
 $z_{\rm h}$ &\,kpc&4 & 6.78 \\ 
 $D_0$ & $\mathrm{cm}^2\,\mathrm{s}^{-1}$ &4.3$\times 10^{28}$& 7.48$\times 10^{28}$  \\
 $\delta^a$ & &0.415 & 0.361 \\
$V_{\rm{alf}}$ & km$\,\mathrm{s}^{-1}$ &30 & 23.8 \\
$V_{\rm{conv}}$(z=\,0\,kpc) & km$\,\mathrm{s}^{-1}$ &0 & 26.9 \\
$\mathrm{d}V_{\rm{conv}}/\mathrm{d}z$ & km$\,\mathrm{s}^{-1}\,\mathrm{kpc}^{-1}$ & 9.8& 0  \\ [1ex] 
\end{tabular}
\end{ruledtabular}
\end{table}

\section{ Antinuclei Production in the Galaxy
\label{s-sources}}

The first step of our study consists in the evaluation of the different antideuteron sources in the Galaxy. We consider here three main components: one stemming from the collisions of cosmic rays with the interstellar medium, one due to WIMP dark-matter annihilations, and one from primordial black hole evaporation. The production of antideuterons is studied and interpreted at accelerator-based experiments by means of statistical hadronisation or coalescence models. In the first approach, particles are produced from a fireball at thermal equilibrium with temperatures close to $156$ MeV for collisions at the LHC and their abundance is fixed at the chemical freeze-out, when the inelastic collisions cease~\cite{Andronic:2010qu}.
In this scenario it is improbable that 'fragile' objects as deuterons (binding energy $= 2.2$ MeV) can survive the hot system created at RHIC or LHC, but the measured (anti)nuclei yields are reproduced by the existing models~\cite{Braun-Munzinger:2018hat}.
It is clear however, that such models do not provide any detailed information on momentum distribution of the formed antideuterons and are also not suited to extract a general formalism for the production of (anti)nuclei in hadron--hadron collisions at energies between 17 GeV (NN threshold for antideuteron production) and several TeV. This very broad energy range characterises the production of antideuterons in our Galaxy by collisions of cosmic rays with the nuclei in the interstellar medium.
Coalescence models, on the other hand, provide predictions for both the yields and momentum spectra of the produced light antinuclei and can be applied to the full energy range. Thus, in this work we consider only coalescence models.

\subsection{Antinuclei formation with coalescence models\label{sec:coalescence}}

There is no first-principle calculation of antideuteron production cross sections in low-energy proton-proton collisions or hypothetical dark-matter annihilation processes. Therefore, any prediction of antideuteron fluxes needs to rely on experimental data.
As antideuteron production is a rare process, experimental data are scarce, and a purely data driven approach is unfeasible.
Instead one has to rely on physically-motivated models of antinuclei coalescence to extrapolate the available data.
The degree of confidence in the final result then depends both on the experimental data, as well as on the plausibility of the model.

The separation of energy scales justifies treating antideuteron production as the coalescence of two antinucleons produced in some high-energy process.
Under the assumption that antineutrons and antiprotons are produced uncorrelated, the antideuteron yield of a process can be estimated by \emph{factorised coalescence}~\cite{Butler:1963pp,Duperray:2005si} as
\begin{eqnarray}
  E_{\overline{\rm d}}\frac{\text d^3N_{\overline{\rm d}}}{\text d p^3_{\overline{\rm d}}}
  \simeq &&B_2\left(E_{\overline{\rm p}}\frac{\text d^3N_{\overline{\rm p}}}{\text d p^3_{\overline{\rm p}}}\right)\left(E_{\overline{\rm n}}\frac{\text d^3N_{\overline{\rm n}}}{\text d p^3_{\overline{\rm n}}}\right)\\
  &&\simeq B_2\left(E_{\overline{\rm p}}\frac{\text d^3N_{\overline{\rm p}}}{\text d p^3_{\overline{\rm p}}}\right)^2,
\end{eqnarray}
where $p_i$ and $\text dN_i/\text dp_i$ are the momentum and the differential yield of particle $i$.
Here, $B_2$ is the coalescence parameter, which can be related to a momentum-space coalescence condition: a pair of anti\-nucleons produced in the same high-energy process coalesces into an antideuteron if the anti\-nucleons' relative momentum in the pair's center-of-mass frame is less than the \emph{coalescence momentum}, ${\left| \vec{p}_{\pbar} - \vec{p}_{\nbar} \right| < p_0}$.
In the analytical factorised coalescence model, this is equivalent to~\cite{Kadastik:2009ts,PhysRevD.99.023016}

\begin{equation} \label{eqn:p0}
B_2 = \frac{1}{8} \frac{4\pi p_0^3}{3} \frac{m_{\overline{\rm d}}}{m_{\overline{\rm p}}^2}\,.
\end{equation}
Since the $B_2$ parameter can be measured~\cite{B2measured}, it is possible to determine the $p_0$ parameter, which will vary as a function of the colliding system and colliding energy. 

\begin{figure}[t]
\includegraphics[width=0.48\textwidth]{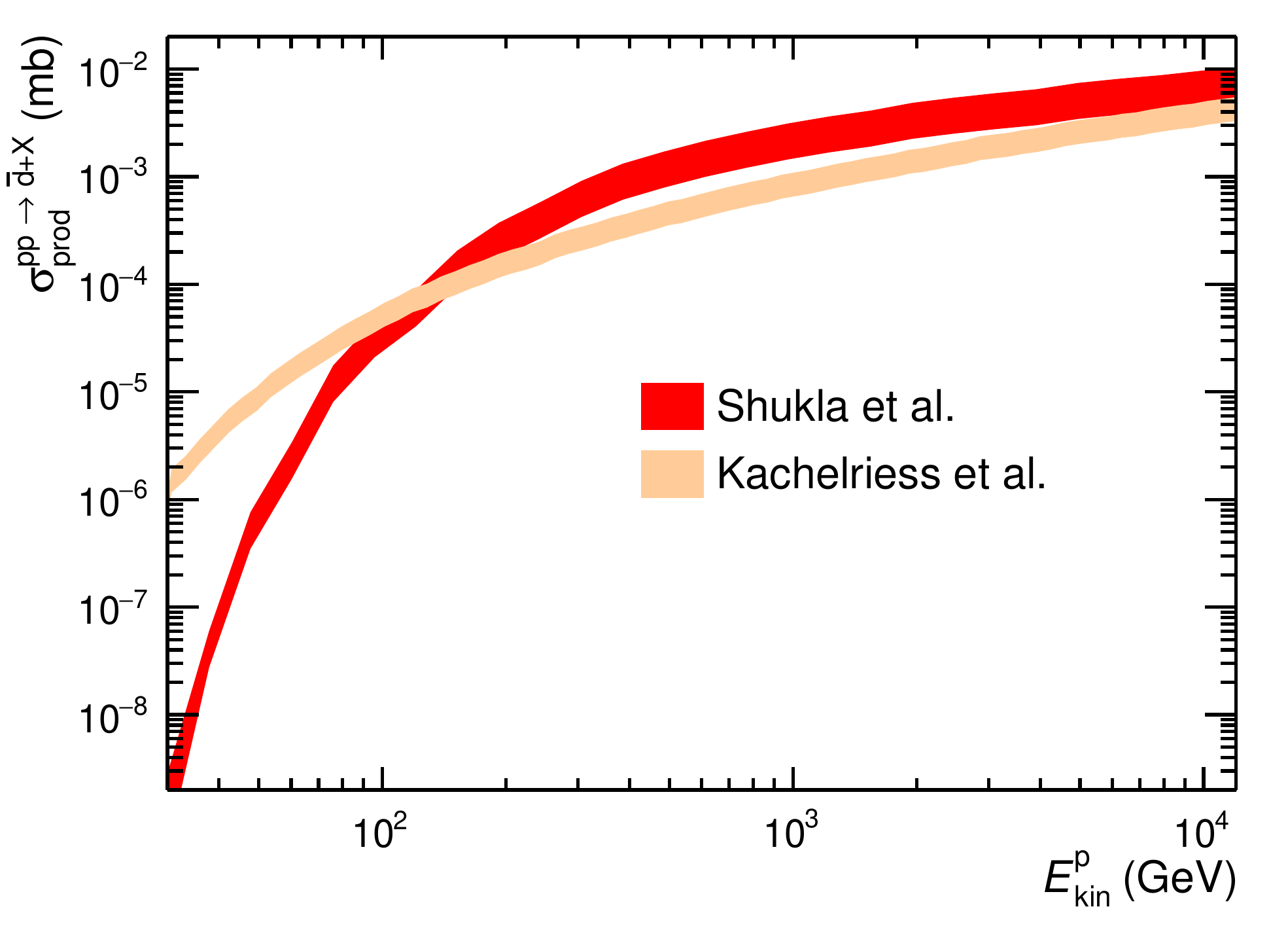}
\caption{Antideuteron total production cross section, in p--p collisions, as a function of the projectile kinetic energy $E^{\rm p}_{\rm kin}$ (GeV) in the laboratory frame for two models~\cite{Gomez-Coral:2018yuk,shukla_prd} and \cite{Kachelriess2020}. The band width corresponds to the uncertainty of the coalescence parameter. See text for details.}
\label{fig:dbar_Xsec}
\end{figure}

Going beyond the factorised coalescence model, an event-by-event procedure based on the Monte Carlo production of antinucleons in particle collisions followed by the coalescence of antiproton--antineutron pairs takes into account correlations between the antinucleons~\cite{Kadastik:2009ts,Ibarra2013a,Ibarra:2012cc,Fornengo:2013osa}.
Here, in addition to the momentum-space coalescence criterion, the coalescing antinucleons are required to be close in position space, excluding antinucleons produced in weak decays from coalescing with those produced promptly.
The coalescence momentum is then determined by comparison to experimental data, where it is found that different values are needed to accommodate results in different particle physics processes~\cite{Aramaki:2015pii}, or even at different process energies~\cite{Gomez-Coral:2018yuk}. The degree of confidence in the final antideuteron yield then depends both on the event generator's ability to accurately simulate antinucleon production and correlation, as well as on the coalescence condition.
 The simplifying assumption of uncorrelated production of antiproton and antineutron in the factorised coalescence model has clear limits, and different experimental results require vastly different values of $B_2$ and $p_0$~\cite{Kadastik:2009ts,Gomez-Coral:2018yuk}. 

A more advanced parameterization of the dependence between $B_2$ and $p_0$ has been investigated by including the size of the formed antinuclei relative to the size of the particle-emitting source formed after the hadron collisions~\cite{Bellini:2018epz}. The size effect is found to be more important for large colliding systems such as Pb--Pb. However, since the size of the particle emitting source depends on the antinuclei transverse momentum~\cite{ALICE:2020ibs}, the interplay between the nuclei size and source size should ideally be considered also in p--p collisions.
More sophisticated approaches in this direction~\cite{Blum:2019suo,Bellini:2020cbj} study the connection between final-state interactions and resulting two-particle correlations for antinucleons and coalescence models. The derived correlation--coalescence relation takes into account both the antinuclei wave functions and colliding system's properties and the size of the particle emitting source.
Recently, a Wigner-function based, semi-classical model 
has been developed~\cite{Kachelriess2020}. Given an ansatz for the antideuteron wave function, the antideuteron yield depends on the spatial spread $\sigma$ of the produced antinucleons, which is fit to antideuteron data. 
In this model,
a single value $\sigma \simeq 1 \,\mathrm{fm}$ explains antideuteron data from different processes and energies~\cite{Kachelriess:2020uoh}.

 In this work, we use the antideuteron production cross sections in p--p collisions determined in~\cite{Gomez-Coral:2018yuk,shukla_prd}, hereafter called Shukla et al., and~\cite{Kachelriess:2020uoh}, hereafter called Kachelrie{\ss} et al. In Shukla et al.~the formation of (anti)deuterons has been studied using multiple MC event generators, settling on \textsc{epos-lhc}~\cite{Pierog:2013ria} for its consistency with antiproton-production data in a wide range of energies. The $p_0$ parameterization for antideuteron production using \textsc{epos-lhc} has been found to depend on the collision energy, initially growing rapidly after the antideuteron production threshold, and finally reaching a saturation value of $p_0/2 = 89.6\,\MeV$ at high energies. Fig.~\ref{fig:dbar_Xsec} shows the total production cross section for antideuterons as a function of collision kinetic energy predicted by these two models. In both cases the bands represent the uncertainty derived from the limited knowledge of the coalescence parameters values ($p_0$ or $\sigma$) after being fitted to data. These uncertainties range from 10\% for Kachelrie{\ss} et al.~to 30\% for Shukla et al. The two approaches use different event generators and consider slightly different data sets, which results in different final cross sections. At low energy ($<$100\,GeV/nucleon), data from Serpukhov\,\cite{Abramov_Baldin1987} was included only in the study by Shukla et al. This dataset showed a lower antideuteron yield than was expected from observations at higher energies~\cite{Gomez-Coral:2018yuk}. At high energies ($>$200\,GeV/nucleon), $p_0$ in Shukla et al. is obtained by including data on antideuteron and antihelium production up to $\sqrt{s}= \text{7 and 13\,TeV}$~\cite{Gomez-Coral:2018yuk,shukla_prd}. The comparatively smaller results obtained by Kachelriess et al.\ around 1000 GeV may be related to the underproduction of antinucleons in QGSJET-II at those energies, with associated uncertainties not quantified by~\cite{Kachelriess:2020uoh}. More measurements of antiproton and antideuteron production cross sections in p-p and p-He collisions are necessary to tune event generators and improve antinuclei formation models.

For the antideuteron production in dark-matter annihilation and primordial black hole evaporation, we employ results from~\cite{Ibarra:2012cc,Herms:2016vop}, to which we refer for a detailed discussion.
These  were determined using the event generator \textsc{Pythia} 8.176 \cite{Sjostrand:2007gs}. There are no data on antideuteron spectra from hypothetical dark-matter annihilation to fix the coalescence momentum $p_0$, which is instead determined from data on Standard Model processes expected to hadronize similarly (i.e.\ non-hadronic initial states producing electroweak gauge bosons or quark--antiquark pairs).
In this spirit, the central value determined from \textsc{Aleph} data on the antideuteron yield of $Z$-decays~\cite{ALEPH:2006qoi} has been used, $p_0 = 192 \MeV$, with an uncertainty bracketed by the $2\sigma$-allowed values determined from \textsc{Aleph} and \textsc{BaBaR} data~\cite{BaBar:2014ssg} (the latter on antideuteron production from $e^+e^-$ collisions at $\sqrt{s}=10.58 \GeV$): $p_0 =128.7 - 226.1 \MeV$~\cite{Ibarra:2012cc}.

\subsection{Cosmic-Ray Collisions with Interstellar Medium}
\begin{figure}
    \centering
    \includegraphics[width=0.48\textwidth]{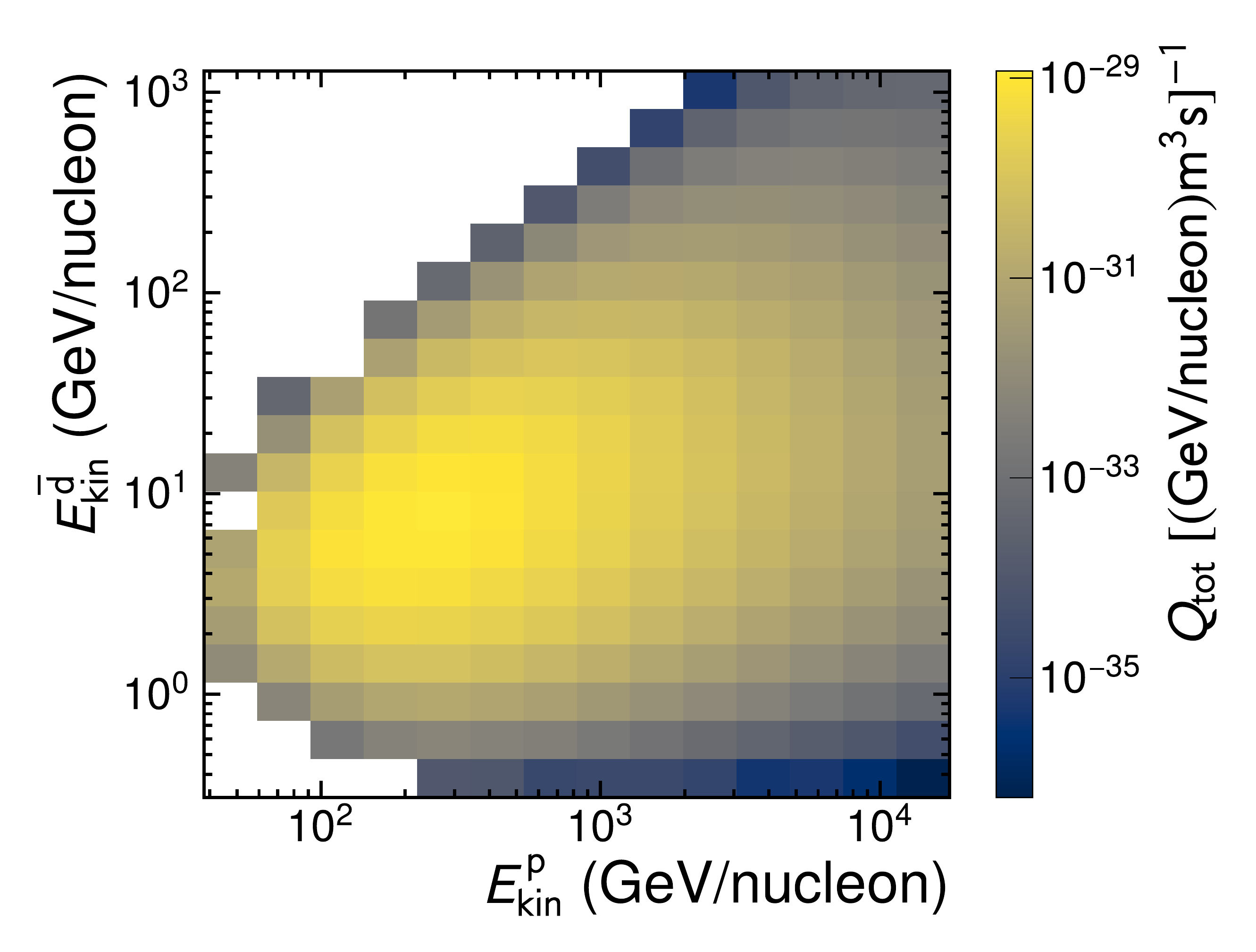}
    \includegraphics[width=0.45\textwidth]{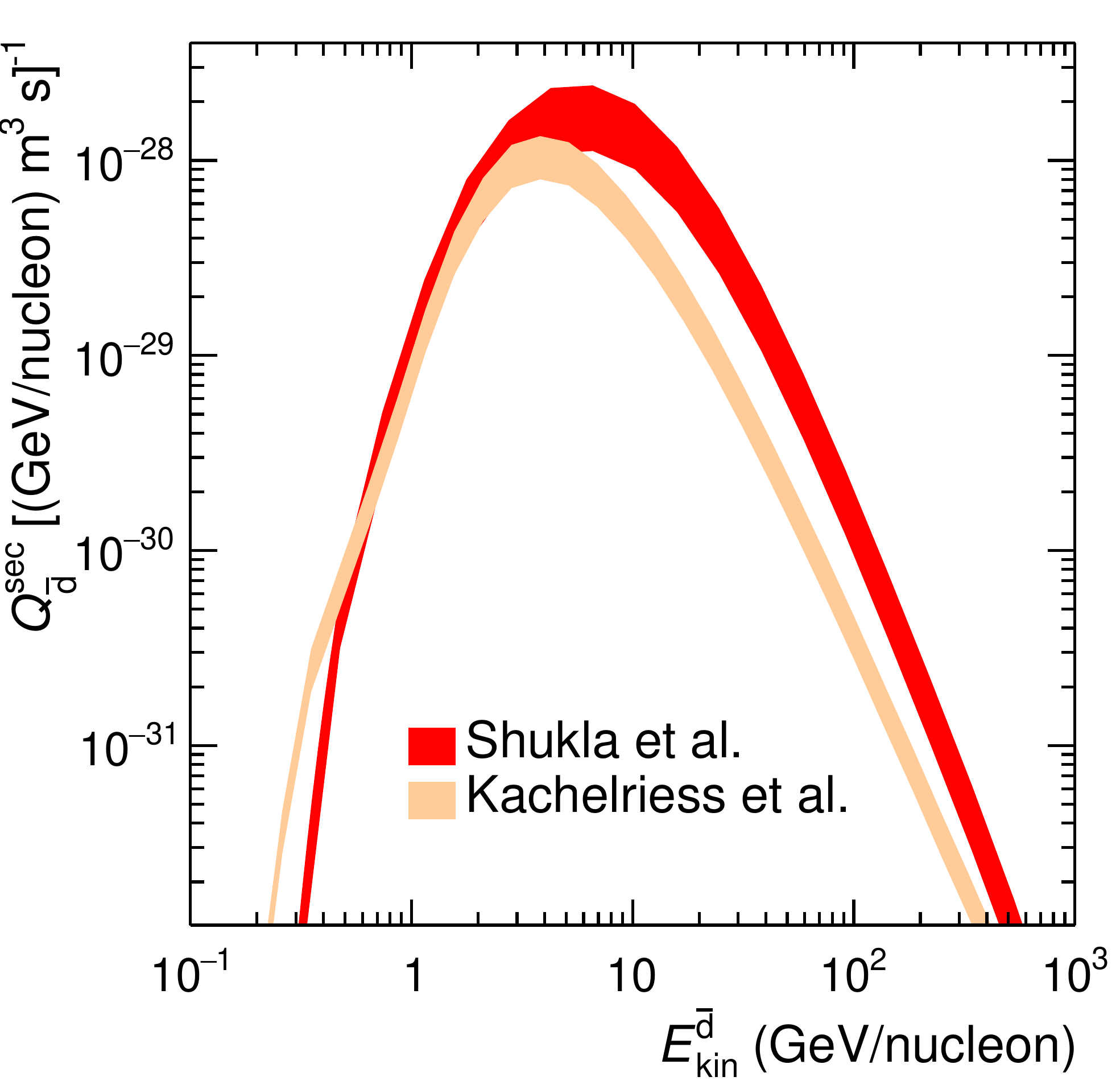}
    \caption{(Upper panel) Contributions to the local source term $Q_{\bar{\rm d}}^{\mathrm{sec}}(E^{\bar{\rm d}}_{\rm{kin}})$ by incoming proton energy $E^{\rm p}_\mathrm{kin}$.The source term includes only p-H collisions. (Lower panel) Antideuteron source term, integrated over proton, helium and antiproton fluxes, as a function of its kinetic energy per nucleon for the two models described in Sec.\ref{sec:coalescence}. The source term on the lower panel includes antideuterons produced in p--H, p--He, He--p, $\bar{\mathrm{p}}$--p, and $\bar{\mathrm{p}}$--He collisions.}
    \label{fig:dbar_source}
\end{figure}

 As described above, cosmic-ray antinuclei are expected to be produced by the interaction of primary cosmic rays, mostly protons and helium nuclei, with the ISM, also composed primarily of hydrogen and helium. Cosmic-ray antideuterons can be formed when the center-of-mass energy of the nucleon--nucleon collision induced by cosmic rays is above an energy threshold of $\sqrt{s} \approx 6$ GeV. Such antideuterons constitute the secondary source term. Although the antideuteron production cross section increases with the collision energy, the steeply declining cosmic-ray proton spectrum causes the antideuteron production to decrease at high energies. The contributions to the antideuteron source term from cosmic-ray protons of different kinetic energies is shown in the upper panel of Fig.~\ref{fig:dbar_source} for p-H collisions. One can see that the largest contribution to the antideuteron yield comes from cosmic-ray energies around 300 GeV.
 The secondary source term $Q_{\bar{\text{d}}}^{sec}$ to be included in the transport Eq.~\ref{eqn:TransportEquation} is calculated using
\begin{widetext}
\begin{equation}\label{eqn:dbarsource}
Q_{\bar{\rm d}}^{\mathrm{sec}}(\textbf{\textit{r}},E^{\bar{\rm d}}_{\rm kin}) = \displaystyle\sum_{i=\text{p}, \text{He}, \bar{\text{p}}} \displaystyle\sum_{j=\text{p}, \text{He}} 4\pi n_j(\textbf{\textit{r}}) \int\limits_{E^{i}_{\rm kin,min}}^\infty \mathrm{d}E^{i}_{\rm kin} \left ( \frac{\text d\sigma_{\rm prod}}{\text dE^{\bar{\rm d}}_{\rm kin}} \right )_{ij} \Phi_i(\textbf{\textit{r}}, E^{i}_{\rm kin}).
\end{equation}
\end{widetext}

Here the index $i$ represents all the incident cosmic-ray species with flux $\Phi_i$ and kinetic energy per nucleon $E^{i}_{\rm kin}$. Index $j$ represents the ISM components with number densities $n_{\rm p}=\text{0.9 }\text{cm}^{-3}$ and $n_{\rm{He}}=\text{0.1 }\text{cm}^{-3}$ used to calculate source functions shown on the lower panel of  Fig.~\ref{fig:dbar_source}. When using the \textsc{galprop} code, the standard implementation of gas distributions in the Galaxy is used, which is based on the available $\mathrm{H} \mathrm{I}$ and $\mathrm{CO}$ surveys as well as the information on ionized compontent~\cite{Moskalenko:2001ya}. The secondary source term convolves the antideuteron differential production cross section $(\text d\sigma_{\rm prod}/\text dE^{\bar{\text{d}}}_{\rm kin})_{ij}$ with the primary cosmic-ray fluxes involved in the collision. The source term is  used in Eq.~\ref{eqn:TransportEquation} to calculate the propagated secondary antideuteron flux using \textsc{galprop}.

In Ref.~\cite{shukla_prd}, $Q_{\bar{\text{d}}}^{\mathrm{sec}}$ was estimated by simulating p--p interactions using \textsc{epos-lhc} as part of the \textsc{crmc}~\cite{crmcRef} package at 27 logarithmically-spaced collision energies between 31\,GeV and 12.5\,TeV in the laboratory frame. Since p--p collisions contribute  60\%-70\% of the total antinuclei source terms~\cite{GomezCoral:2673048, Korsmeier:2017xzj}, only those have been simulated. The $\mathrm{p}$--He, He--$\mathrm{p}$, and He--He contributions have been estimated by scaling the parameterization developed in Ref.~\cite{Reinert:2017aga}. The differential production cross section has to be implemented in \textsc{galprop} in logarithmic kinetic energy per nucleon bins for protons and antideuterons. For this purpose the results from~\cite{shukla_prd} have been interpolated using a cubic polynomial. An additional contribution to the antideuteron source term, especially important at low energies, is the interaction of cosmic-ray antiprotons with the ISM. This contribution was taken into account by simulating antideuteron production in $\overline{\text{p}}$--p and $\overline{\text{p}}$--He collisions using \textsc{epos-lhc}~\cite{GomezCoral:2673048}. In Ref.~\cite{Kachelriess:2020uoh}, the antideuteron production in all collision systems (p--H, p--He, He--p, $\bar{\mathrm{p}}$--p, and $\bar{\mathrm{p}}$--He) was estimated using QGSJET-II. The lower panel of~\Fig{fig:dbar_source} shows the resulting secondary antideuteron source term, integrated over proton, helium and antiproton fluxes, as a function of the kinetic energy per nucleon for Shukla et al.~\cite{Gomez-Coral:2018yuk,shukla_prd} and Kachelrie{\ss} et al.~\cite{Kachelriess:2020uoh}. The discrepancies observed between the two source terms below 0.5~GeV/nucleon and above 3~GeV/nucleon are a consequence of the disagreement in cross section described in the last section. The width of the two bands corresponds to the uncertainty in the coalescence model (see Sec.~\ref{sec:coalescence}).

\subsection{Dark Matter\label{sec-DMsources}}
There is compelling evidence from multiple astronomical and cosmological observations for the presence of large amounts of dark matter in the Universe and our Galaxy (e.g.~\cite{Bertone:2004pz}). If dark matter consists of particles that can annihilate into Standard Model (SM) particles, as is generically expected for WIMPs produced through thermal freeze-out in the early Universe, it constitutes a potential source of cosmic-ray antideuterons.
The source term for cosmic-ray antideuterons from dark-matter annihilations is 
\begin{equation}
 Q^{\mathrm{DM,ann}}_{\bar{\rm d}}(\textbf{\textit{r}},E^{\bar{\rm d}}_{\rm{kin}}) = \frac{1}{2} \left(\frac{\rho(\textbf{\textit{r}})}{m_{{\mathrm{DM}}}}\right)^2 \sigmav_f \frac{\mathop{\text dN_f^{\bar{\rm d}}}}{\mathop{\text dE^{\bar{\rm d}}_{\rm{kin}}}}\,,
 \label{DMSource}
\end{equation}
where $\rho(\vec r)$ is the local dark-matter density in the Galaxy, $\sigmav_f$ is the velocity averaged dark-matter annihilation cross section into channel $f$, for example b$\bar{\rm b}$ or W$^+$W$^-$, and $\mathop{\text dN_f^{\bar{\rm d}}}/\text dE^{\bar{\rm d}}_{\rm{kin}}$ is the antideuteron multiplicity produced from one such annihilation event. These are examined in turn in the following.

\begin{figure}[htbp]
\centering
\includegraphics[width=0.47\textwidth]{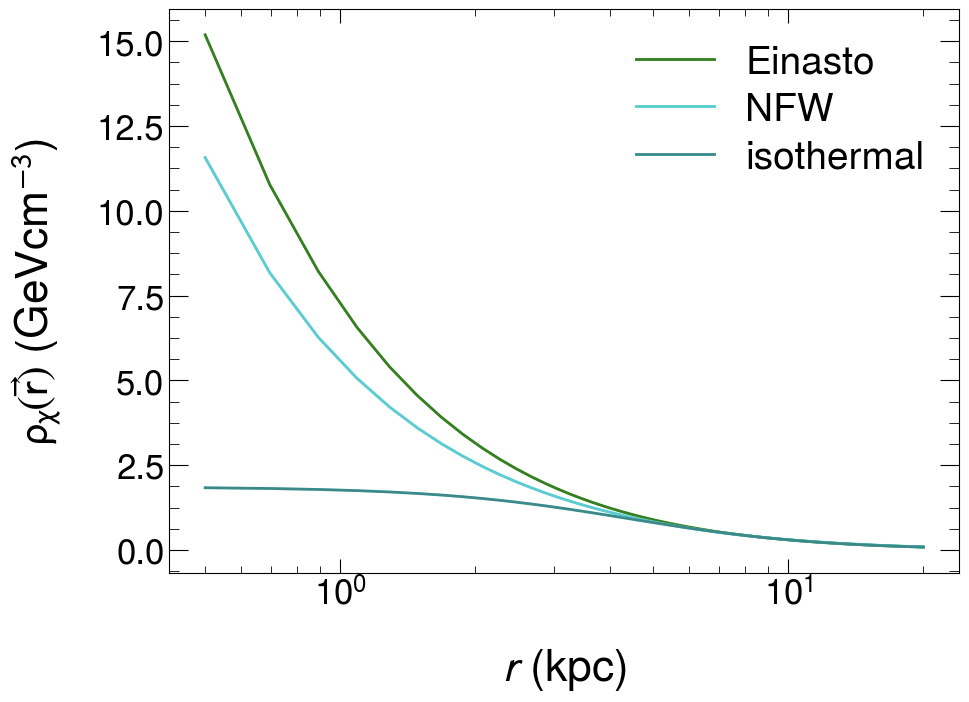}
\caption{The dark-matter density profiles, as a function of distance from the Galactic center.} \label{fig:DM_profiles_log}
\end{figure}

The Galactic dark-matter distribution $\rho(\vec r)$ can be determined from kinematic tracers~\cite{deSalas:2020hbh}. There is uncertainty both in the local density at the position of the Sun $r_\odot$, as well as in the shape of the distribution towards the inner Galaxy. To bracket the uncertainty due to the dark-matter profile, we consider the cuspy Navarro-Frenk-White~\cite{Navarro:1996gj} profile $\rho_\mathrm{NFW}(\vec{r}) \propto (r/r_s)^{-1} (1+r/r_s)^{-2}$ with scale radius $r_s = 24.4\,\mathrm{kpc}$. We also consider the flatter isothermal profile ~\cite{Begeman:1991iy} $\rho_\mathrm{isothermal}(\vec{r}) \propto (r^2 + r_s^2)^{-1}$ with scale radius $r_s = 4.38$ kpc, as well as the very cored Einasto profile~\cite{Ibarra:2012cc} $\rho_\mathrm{Einasto} \propto \exp \left( -\frac{2}{\alpha} \left[ (r/r_s)^\alpha -1\right]\right)$ with scale radius $r_s = 28.44$ and $\alpha = 0.17$.
These are normalised to a local dark-matter mass density of $\rho_\odot = 0.4 \GeV/\text{cm}^3$, which is uncertain by up to a factor of 2~\cite{Benito:2020lgu}. The profiles are shown in Fig.~\ref{fig:DM_profiles_log}.

The value of the dark-matter annihilation cross section is not known.
In the freeze-out scenario of particle dark-matter production, the dark-matter relic abundance today predicts a value $\sigmav_\mathrm{thermal} \sim %1 \,\mathrm{pb}\, c = 3
2 \times 10^{-26} \mathrm{cm}^3/\mathrm{s}$ in the early Universe (e.g.~\cite{Bringmann:2020mgx}). In this context, the dark-matter mass $m_\mathrm{DM}$ is expected to be in the GeV--TeV range, with lower masses typically affected by constraints on dark-matter annihilation during recombination~\cite{Planck:2015fie} and larger masses in conflict with unitarity~\cite{Griest:1989wd}.
To be relevant in the context of cosmic-ray antideuterons, the dark-matter mass needs to lie in the GeV range, with smaller masses unable to produce antideuterons, while the overall annihilation rate drops as $m_\mathrm{DM}^{-2}$.

%{\footnotesize  Two channels were analyzed: i) dark matter annihilating into $\bar{\Lambda}_b$ directly and ii) dark matter annihilating into 14 GeV light mediators which then decay into $\bar{\Lambda}_b$. These were also analysed using the event generator \textsc{Pythia} 8.XXX, in particular by tuning \textsc{Pythia} to reproduce rates of $\bar{\Lambda}_b$ production as were observed by LEP \cite{} {\color{blue}(see however~\cite{Kachelriess:2021vrh,Winkler:2021cmt})}. It was shown that these decays via $\bar{\Lambda}_b$ can boost the momentum integrated flux of antideuterons from dark matter annihilations by a factor of up to 15\%, and in the light mediator channel shift the peak of the produced antideuteron spectra to far higher energies. \jc{instead of the following, use `see howewer' above? I'm not sure the $\bar{\Lambda}_b$ yields are Kachelriess' issue, thought it was $p,\pi,K$ yields in $\bar{\Lambda}_b$-tune.} In this discussion, one has to take into account that the tuned $\bar{\Lambda}_b$ yields in \cite{Winkler:2020ltd} are not in agreement with complementary yields measured at colliders \cite{Kachelriess:2021vrh}. }

Motivated by reports of tentative  $^3\overline{\mathrm{He}}$ candidate events at relative high energies by the AMS-02 collaboration \cite{antihe, antihe2, antihe3}, other decays channels were recently considered.
$\bar{\Lambda}_{\rm b}$ decays into antinuclei have received increasing attention, and it has been shown that they may increase the flux of antideuterons from dark matter by up to a factor of 4 \cite{Winkler:2020ltd} (see however~\cite{Kachelriess:2021vrh,Winkler:2021cmt}).

In the following, results for several benchmarks are shown intended to illustrate a range of antideuteron fluxes that can realistically be expected. The spectra are determined by dark-matter mass and annihilation channel, for which we choose $m_{\rm{DM, b\bar b}} = 10, 51, 100, 1000 \GeV$ and $m_{\rm{DM, WW}} = 94, 100, 1000 \GeV$.
The normalization is essentially determined by the annihilation cross section \sigmav. There are strong constraints as well as contentious signal hints~\cite{Cholis:2019ejx,Reinert:2017aga,Cui:2018klo,Heisig:2020nse} on \sigmav\ from antiproton measurements by the AMS-02 detector~\cite{AMS:2016oqu}. These however depend significantly on the modeling of the antiproton production cross section, the propagation model, as well as experimental uncertainties that are yet to be fully characterised by the AMS-02 collaboration.
We choose a conservative approach, adopting the theoretical expectation of $\sigmav_\mathrm{thermal} \sim 2 \times 10^{-26} \mathrm{cm}^3/\mathrm{s}$, which is not conclusively ruled out by antiproton measurements for dark-matter masses between $\sim 50 - 100 \GeV$, while for the $10 \GeV$ benchmark we adopt $\sigmav = 3 \times 10^{-27} \mathrm{cm}^3/\mathrm{s}$.
These values are compatible with gamma ray limits from dwarf spheroidal galaxies~\cite{Fermi-LAT:2016uux}.
Since $\sigmav$ only affects the normalisation, our results are easily translated to different values. We also propagate injection spectra determined by~\cite{Winkler:2020ltd} with enhanced antideuteron production in $\bar{\Lambda}_{\rm b}$ decay: i) $67\GeV$ dark matter annihilating into $\bar{\Lambda}_{\rm b}$ directly and ii) $80\GeV$ dark matter annihilating into 14 GeV light mediators which then decay into $\bar{\Lambda}_{\rm b}$.

\subsection{Primordial Black Holes}

Primordial Black Holes (PBHs) could have been formed in the early Universe~\cite{1971MNRAS.152...75H,1967SvA....10..602Z} and constitute a fraction of the cold dark matter abundance today. Their signatures today depend crucially on their mass. If they are sufficiently light, $M\lesssim 5\cdot10^{14}\,\mathrm{g}$, they are predicted to evaporate through Hawking radiation~\cite{1974Natur.248...30H,1975CMaPh..43..199H} on the timescale of the age of the Universe. The energy scale of the emitted particles is given by the Hawking temperature $T \approx 1.06 \GeV/\left(M_\mathrm{PBH} / 10^{13}\,\mathrm{g}\right)$, evidently sufficient to produce antinuclei for such light PBHs.

Antideuteron source spectra from PBH evaporation have been computed in the event-by-event coalescence model as described in~\cite{Herms:2016vop}.
These are obtained by integrating the instantaneous Hawking emission rate of a single PBH over the PBH mass distribution today.
While the initial mass function of PBHs produced in the early Universe is model dependent, the mass spectrum of PBHs capable of producing antideuterons today is determined solely by the mass loss rate. The predicted antinuclei spectra $Q(E^{\bar{\rm N}}_{\rm{kin}})$ are hence completely fixed by Hawking evaporation, and normalization is the only free parameter.

The normalization of the antideuteron source term $Q_\mathrm{PBH}(T,\vec r)$ from PBH evaporation is linked to the PBH number density.
PBHs constitute a form of cold dark matter, and one can assume their number density in the Galaxy follows that of DM, $Q_\mathrm{PBH}(E^{\bar{\rm d}}_{\rm{kin}},\vec r) \propto Q_\mathrm{PBH}(E^{\bar{\rm d}}_{\rm{kin}}) \cdot \rho_\mathrm{DM}(\vec r)$.
Assuming a particular initial PBH mass function, $\mathop{\text dN/\text dM} \propto M^{-5/2}$~\cite{carr_primordial_1975}, the overall normalisation is fixed by the local PBH mass density (or equivalently, the fraction of dark-matter in form of these PBHs). The results for $\rho_\mathrm{PBH} = 4 \cdot 10^{-11} \rho_\mathrm{DM}$ are shown here, which was found to be marginally compatible with antiproton limits in~\cite{Herms:2016vop} \footnote{
As the remaining lifetime of a BH with GeV temperature is only $\sim 10^4 \,\mathrm{yr}$, the population of antinuclei-emitting PBHs can also be characterised by the local rate of explosive final PBH evaporation events. Our antinuclei spectra correspond to a local explosion rate of $3\cdot10^{-4} \,\mathrm{pc}^{-3}\mathrm{yr}^{-1}$.
}.

\section{Antideuteron Inelastic Cross Section\label{s-cross}}

After antideuterons are formed, their inelastic interactions with the ISM lead to a reduction of their flux, so the determination of the corresponding cross sections is a crucial aspect in flux calculations. The probability of an inelastic interaction is determined by the total nuclear inelastic cross section \sigmainel, which includes all processes leading to the disappearance of antideuterons (such as annihilation, nuclear breakup, charge exchange etc.). 

The measurement of the inelastic cross section typically requires a beam of particles of interest (with well-determined momentum) and a target of known material composition and thickness. Since it is very challenging to obtain a beam of antideuterons with precise momentum, the knowledge of \sigmainel\ was until recently very limited. For nearly 50 years the only available measurements of \sigmainel\ came from the experimental facilities at the U-70 proton synchrotron. There, \sigmainel\ was measured on various material targets (Li, C, Al, Cu and Pb) for antideuterons with momenta of 13.3 GeV/$c$~\cite{Denisov:1971im} and 25 GeV/$c$~\cite{Binon:1970yu}.

%\begin{figure}[ht]
%\centering
%\includegraphics[width=0.5\textwidth]{figures/sigma_inel_133GeV.pdf}
%\caption{\sigmainel\ at $p = 13.3$ GeV/$c$ on various target materials as a %function of atomic mass number $A$~\cite{Denisov:1971im}.}
%\label{fig:sigmainel_133gev}
%\end{figure}

\begin{figure}[ht]
\centering
\includegraphics[width=0.48\textwidth]{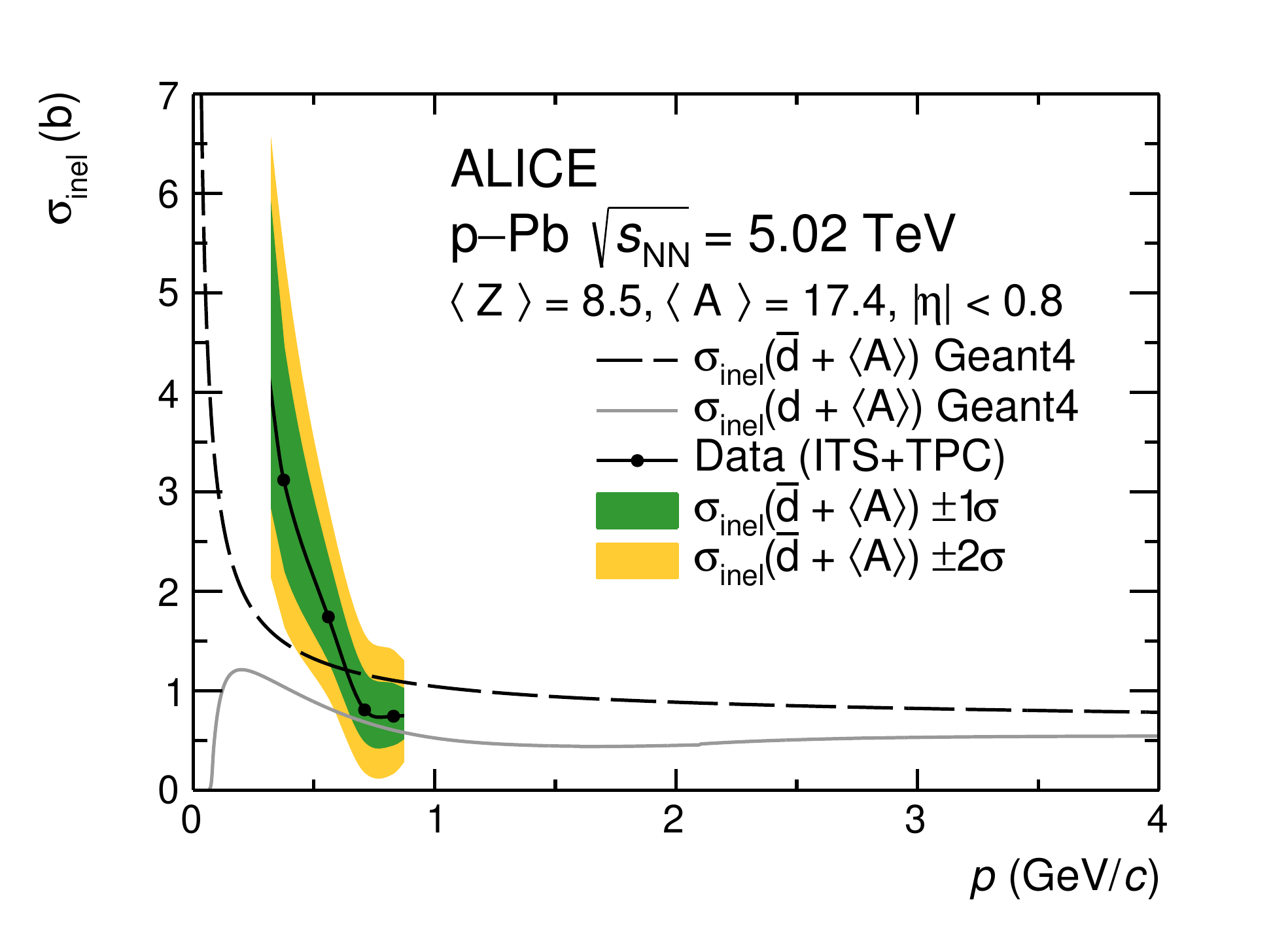}
\includegraphics[width=0.48\textwidth]{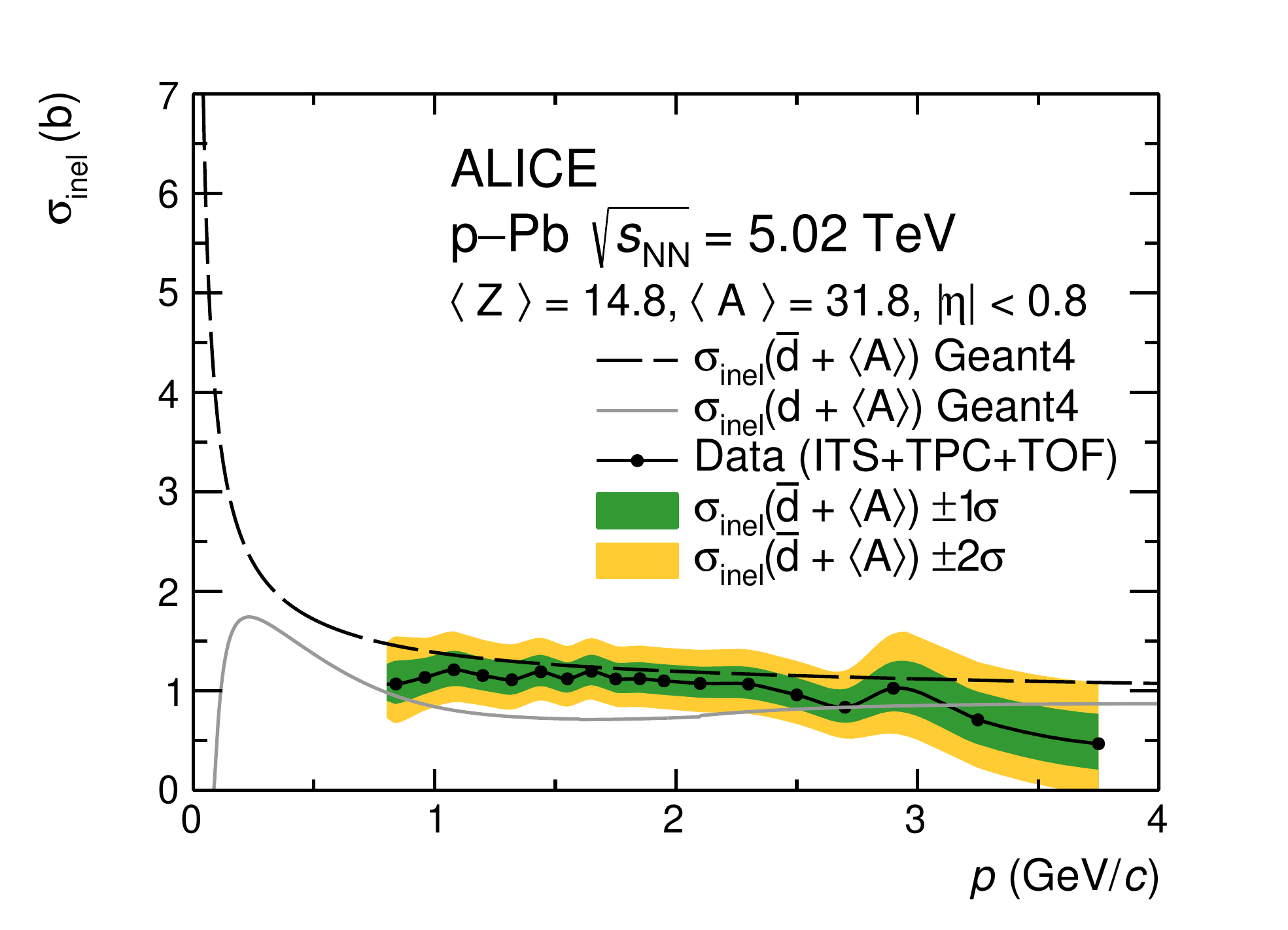}
\caption{\sigmainel\ measured on an average material element of the ALICE detector as a function of the momentum~\cite{ALICE:2020zhb}. Dashed black lines and full gray lines represent the \geant parameterizations for \sigmainel\ and $\sigma_{\rm{inel}}({\rm d})$ respectively. The experimental data points are connected by solid black lines, and green and orange bands correspond to $\pm 1$ and $\pm 2\sigma$ uncertainties on \sigmainel.}
\label{fig:sigmainel_alice}
\end{figure}

In high-energy collisions between protons and lead nuclei at TeV energies, matter and antimatter are abundantly produced in essentially equal amounts~\cite{ALICE:2013yba,ALICE:2010hjm,ALICE:2019fee,ALICE:2017jmf,ALICE:2019dgz,ALICE:2019bnp,ALICE:2020foi,ALICE:2017xrp,ALICE:2015wav}. This fact not only facilitates detailed studies of (anti)nuclei production mechanisms~\cite{Bellini:2020cbj,Andronic:2010qu}, but also allows one to investigate the antinuclei inelastic interactions with the detector material. Last year, the ALICE Collaboration presented novel results of \sigmainel\ in the momentum range below 4 GeV/$c$~\cite{ALICE:2020zhb}. The analysis exploited the antimatter-to-matter ratio method, in which the raw reconstructed antideuteron-to-deuteron ratio ($\bar{\rm d}/{\rm d}$) served as an experimental observable, as it is sensitive to the inelastic cross section of the (anti)nuclei entering the ratio. Since $\sigma_{\rm{inel}}({\rm d})$ at low energies is known~\cite{Auce:1996fe,Jaros:1977it}, the antideuteron inelastic cross section \sigmainel\ could be extracted by comparing the experimental results for raw reconstructed $\bar{\rm d}/{\rm d}$ with Monte Carlo simulations in which $\sigma_{\rm{inel}}({\rm d})$ is constrained by the available data. The resulting \sigmainel\ is shown for the atomic mass numbers of $\langle A \rangle = 17.4$ and 31.8 in Fig.~\ref{fig:sigmainel_alice}. These values of $\langle A \rangle$ were  obtained by weighting the contribution from different materials of the ALICE detector with their density times the path length crossed by particles.

The results from ALICE are in  good agreement within uncertainties with the parameterizations of \sigmainel\ implemented in the \geant\ toolkit~\cite{GEANT4:2002zbu}, which is widely used for the propagation of particles through the matter. In this toolkit the description of antinucleus--nucleus inelastic cross sections is based on Glauber calculations. Direct Glauber model simulations during each propagation step in \geant\ would be computationally too expensive, so the antinuclei inelastic cross sections are parameterized as a function of atomic mass number $A$ of the target nucleus as described in \cite{Uzhinsky:2011zz}:

\begin{equation}
    \sigma^{\rm{inel}}_{hA} = \pi R_{A}^{2} \ln\left(1+\frac{A\sigma^{\rm{tot}}_{hN}}{\pi R^{2}_{A}}\right).
\end{equation} 

Here the total (elastic plus inelastic) cross section $\sigma^{\rm{tot}}_{hN}$ of a hadron $h$ ($h$ = $\mathrm{\overline{p}}$, $\mathrm{\overline{d}}$, $^3\mathrm{\overline{\rm He}}$, or $^4\mathrm{\overline{\rm He}}$) interacting with a nucleon $N$ is estimated with Glauber calculations. $A$ is the atomic number of the target nucleus with radius $R_{A}$, which is parameterized as a function of $A$ using $\sigma^{\rm{inel}}_{hA}$ and $\sigma^{\rm{tot}}_{hN}$ calculated using the  Glauber model for given $h$ and $A$.
% \ls{would be good to mention what R is, if it is a model parameter which is fitted, this should still be mentioned}

An alternative parameterization of \sigmainel\ can be obtained from the total deuteron--antiproton cross section $\sigma_{\rm{tot}}^{\rm d {\overline{\rm{p}}}}$ measured in~\cite{Zyla:2020zbs}. By symmetry, it is equal to the total antideuteron--proton cross section $\sigma_{\rm{tot}}^{\overline{\rm d}\rm{p}}$ which, together with the known total and elastic antiproton--proton cross sections~\cite{Zyla:2020zbs}, can be used to estimate the inelastic antideuteron--proton cross section in the following way:

\begin{equation}
    \sigma_{\rm{inel}}^{\overline{\rm d}\rm{p}} \approx \frac{\sigma_{\rm{tot}}^{\overline{\rm d}\rm{p}}}{\sigma_{\rm{tot}}^{\overline{\rm p}\rm{p}}}(\sigma_{\rm{tot}}^{\overline{\rm p}\rm{p}} - \sigma_{\rm{el}}^{\overline{\rm p}\rm{p}}).
\end{equation}

This approach has been used in previous studies estimating the antideuteron fluxes near Earth~\cite{Korsmeier:2017xzj}.

Another assumption which can be employed to estimate the antideuteron inelastic cross section is that it is simply twice as large as the corresponding antiproton inelastic cross section at the same kinetic energy per nucleon $E_{\rm{kin}}^{\bar{\rm p}} = E_{\rm{kin}}^{\bar{\rm d}}/n$~\cite{Ibarra:2012cc,Donato:2008yx}:

\begin{equation}
    \sigma_{\rm{inel}}^{\overline{\rm d}\rm{p}} (E_{\rm{kin}}^{\bar{\rm d}}/n) \approx  2\sigma_{\rm{inel}}^{\overline{\rm p}\rm{p}} (E_{\rm{kin}}^{\bar{\rm p}}).
\end{equation}
% \ls{these equations should be rewritten with kinetic energy variable Ekin as done in above source equations}
The inelastic antiproton--proton cross section can be taken e.g. from~\cite{Tan:1983de} as:

\begin{eqnarray}
    \sigma_{\rm{inel}}^{\overline{\rm p}\rm{p}} (E_{\rm{kin}}^{\bar{\rm p}}) = 24.7(1&&+0.584 {E_{\rm{kin}}^{\bar{\rm p}}}^{-0.115}\\
    &&+0.856 {E_{\rm{kin}}^{\bar{\rm p}}}^{-0.566})\,\rm{mbarn},
\end{eqnarray}

where $E_{\rm{kin}}^{\bar{\rm p}}$ is in units of GeV. For the inelastic cross section of antideuterons colliding with helium nuclei, the $\sigma_{\rm{inel}}^{\overline{\rm d}\rm{p}}$ can be scaled by the geometrical factor of $4^{2/3}$. The described approach has been used in antideuteron cosmic-ray studies presented in~\cite{Ibarra:2012cc}.

\begin{figure}[ht]
\centering
\includegraphics[width=0.48\textwidth]{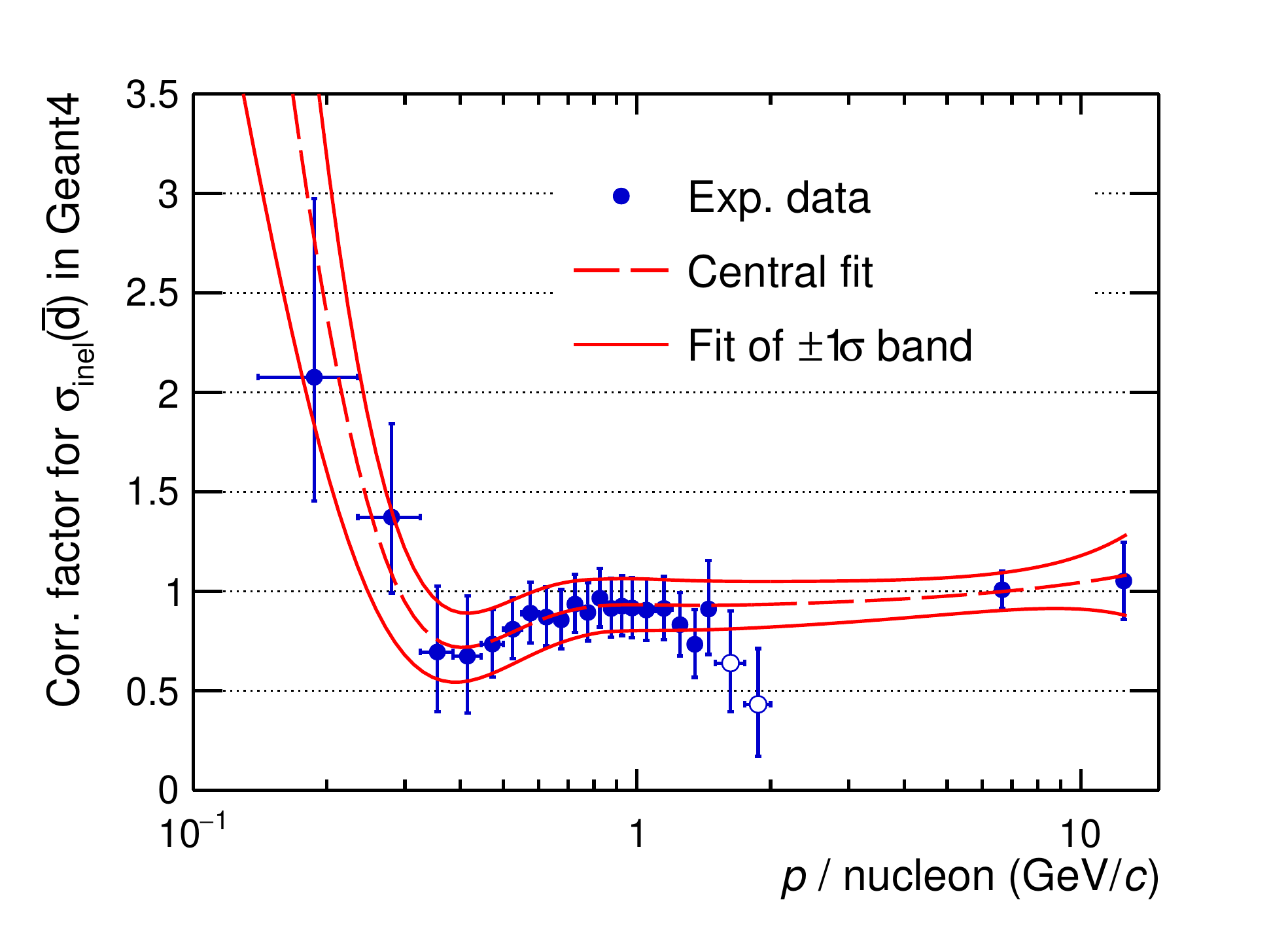}
\caption{Correction factor for \sigmainel\ in \geant\ as a function of the antideuteron momentum per nucleon. The correction factor obtained from experimental data~\cite{ALICE:2020zhb,Denisov:1971im,Binon:1970yu} is shown as blue points, and red lines show the smooth fit used to describe the data. Two data points excluded from the fit are shown as open circles. The numerical values for the correction factor can be found in Table~\ref{tab:inelCS_dbarp}.}
\label{fig:sigmainel_corrfactor}
\end{figure}

Unlike previous estimates, the results presented in the current paper are based entirely on experimental data for \sigmainel. In order to model the inelastic processes of antideuterons with matter, the results published in~\cite{ALICE:2020zhb,Denisov:1971im,Binon:1970yu} are used to obtain a momentum-dependent correction factor for the  \sigmainel\ parameterization implemented in \geant. Fig.~\ref{fig:sigmainel_corrfactor} shows this correction factor as a function of the antideuteron momentum. The experimental data and their uncertainties from~\cite{ALICE:2020zhb,Denisov:1971im,Binon:1970yu} are described with smooth functions using a combination of exponential and polynomial functions in order to interpolate the results into the momentum ranges with no measurements. The last two data points from~\cite{ALICE:2020zhb} at $p / \rm{nucleon} \lesssim 2$ GeV/$c$ have been excluded from the fit to obtain a smooth interpolation between the results from ALICE and from U-70 experiments. For the extrapolation to momenta above the measured momentum range, the correction factor corresponding to the last measured value from~\cite{Binon:1970yu} (at $p / \rm{nucleon} = 12.5$ GeV/$c$) has been considered. The numerical values of the correction factor for \sigmainel\ in \geant\ can be found in Table~\ref{tab:inelCS_dbarp}. %\ls{why correction factor not cross section here?}

The resulting correction factor is applied to \sigmainel\ from \geant\ on all target materials relevant to our studies (mainly hydrogen and helium). Since the experimental data are only available for relatively heavy target elements (e.g. $\langle A \rangle = 17.4$ and $\langle A \rangle = 34.7$ in the case of ALICE results), an additional uncertainty has been assigned to the fit results to take into account a possible dependence of this correction factor on the atomic mass number of target $A$. This uncertainty is taken from the difference between the parameterization for the $A$-dependence implemented in \geant\ and the full Glauber calculation and amounts to $<$ 8\% \cite{Uzhinsky:2011zz}. It is worth to mention that \geant\ parameterizations describe well the antiproton inelastic cross section on various materials~\cite{Uzhinsky:2011zz}. Therefore the deviations of the correction factor from unity seen in Fig.~\ref{fig:sigmainel_corrfactor} are apparently related to the properties of antideuteron and not of the target, i.e.~one should expect a weak dependence of this correction factor on $A$.

\begin{figure}[ht]
\centering
\includegraphics[width=0.48\textwidth]{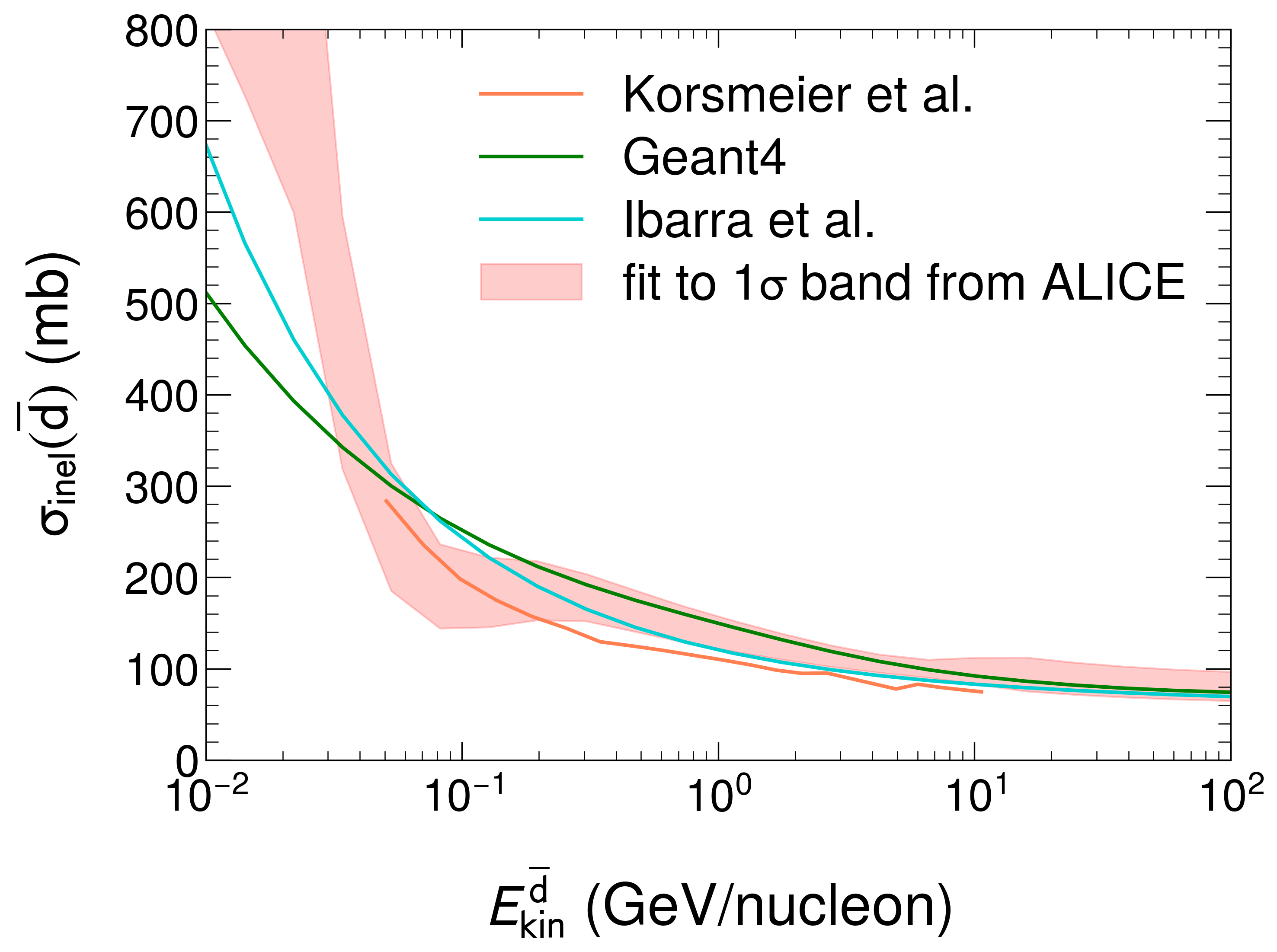}
\caption{Antideuteron--proton inelastic cross section as a function of kinetic energy per nucleon. The light red band shows the fit to available experimental data (see text for details), the dark green line represents the parameterization implemented in \geant, the orange line shows the parameterization employed in \cite{Korsmeier:2017xzj}, and the turquoise line corresponds to the parameterization used in~\cite{Ibarra:2012cc}.}
\label{fig:sigmainel_results}
\end{figure}

The resulting antideuteron--proton inelastic cross section is shown in Fig.~\ref{fig:sigmainel_results} together with the parameterization used in \geant\ and with the models employed in~\cite{Korsmeier:2017xzj} and in~\cite{Ibarra:2012cc}.
\begin{table}
\caption{\label{tab:inelCS_dbarp}%
Correction factor for \sigmainel\ implemented in \geant. Data points enclosed in brackets have not been used for the fit.}
\begin{ruledtabular}
\begin{tabular}{cc}
 \begin{tabular}{@{}c@{}} Antideuteron momentum \\ per nucleon, GeV/$c$ \end{tabular} & \begin{tabular}{@{}c@{}} Correction factor \\ for \sigmainel \end{tabular} \\ [0.5ex] 
\colrule
    \hline
    $0.1875 \pm 0.0475$ & $2.076^{+0.897}_{-0.622}$ \\
    $0.28 \pm 0.045$ & $1.373^{+0.470}_{-0.381}$ \\
    $0.355 \pm 0.03$ & $0.694^{+0.332}_{-0.301}$ \\
    $0.415 \pm 0.03$ & $0.674^{+0.303}_{-0.287}$ \\
    $0.4725 \pm 0.0275$ & $0.735^{+0.175}_{-0.165}$ \\
    $0.525 \pm 0.025$ & $0.809^{+0.156}_{-0.149}$ \\
    $0.575 \pm 0.025$ & $0.890^{+0.156}_{-0.150}$ \\
    $0.625 \pm 0.025$ & $0.870^{+0.151}_{-0.145}$ \\
    $0.675 \pm 0.025$ & $0.856^{+0.153}_{-0.146}$ \\
    $0.725 \pm 0.025$ & $0.935^{+0.150}_{-0.144}$ \\
    $0.775 \pm 0.025$ & $0.893^{+0.148}_{-0.142}$ \\
    $0.825 \pm 0.025$ & $0.965^{+0.150}_{-0.145}$ \\
    $0.875 \pm 0.025$ & $0.914^{+0.150}_{-0.145}$ \\
    $0.925 \pm 0.025$ & $0.925^{+0.153}_{-0.147}$ \\
    $0.975 \pm 0.025$ & $0.916^{+0.154}_{-0.148}$ \\
    $1.05 \pm 0.05$ & $0.906^{+0.157}_{-0.152}$ \\
    $1.15 \pm 0.05$ & $0.914^{+0.163}_{-0.156}$ \\
    $1.25 \pm 0.05$ & $0.832^{+0.162}_{-0.156}$ \\
    $1.35 \pm 0.05$ & $0.733^{+0.173}_{-0.166}$ \\
    $1.45 \pm 0.05$ & $0.911^{+0.244}_{-0.229}$ \\
    $(1.625 \pm 0.125)$ & $(0.639^{+0.261}_{-0.244})$ \\
    $(1.875 \pm 0.125)$ & $(0.431^{+0.281}_{-0.262})$ \\
    $6.65$ & $1.0075 \pm 0.0950$ \\
    $12.5$ & $1.052 \pm 0.193$ \\
\end{tabular}
\end{ruledtabular}
\end{table}

\section{Results}
This section presents predictions for antideuteron fluxes calculated for observations near Earth from the sources discussed in section~\ref{s-sources}. In addition it highlights the different uncertainties --- well quantified and qualitative ones --- entering the calculation and discusses their importance.

The various antideuteron source functions described in section~\ref{s-sources} were implemented in the \textsc{galprop} cosmic-ray propagation code and the inelastic cross sections described in section~\ref{s-cross} were also included in the \textsc{galprop} transport equation (see Eq.~\ref{eqn:TransportEquation}). The propagation parameters considered as a default have been taken from~\cite{2020ApJS..250...27B} and systematic studies of the impact of a different implementation of these parameters have been carried out, also considering the work of~\cite{Cuoco:2019kuu}.
The effect of solar modulation is modeled employing the force-field approximation with constant Fisk potential $\phi=0.5$ GV corresponding to solar minimum. The fluxes are shown after solar modulation only in Fig.~\ref{fig:antifeuteronTOA}.

\subsection{Cosmic-Ray Fluxes}
\begin{figure*}[htbp]
\centering
\includegraphics[width=0.45\textwidth]{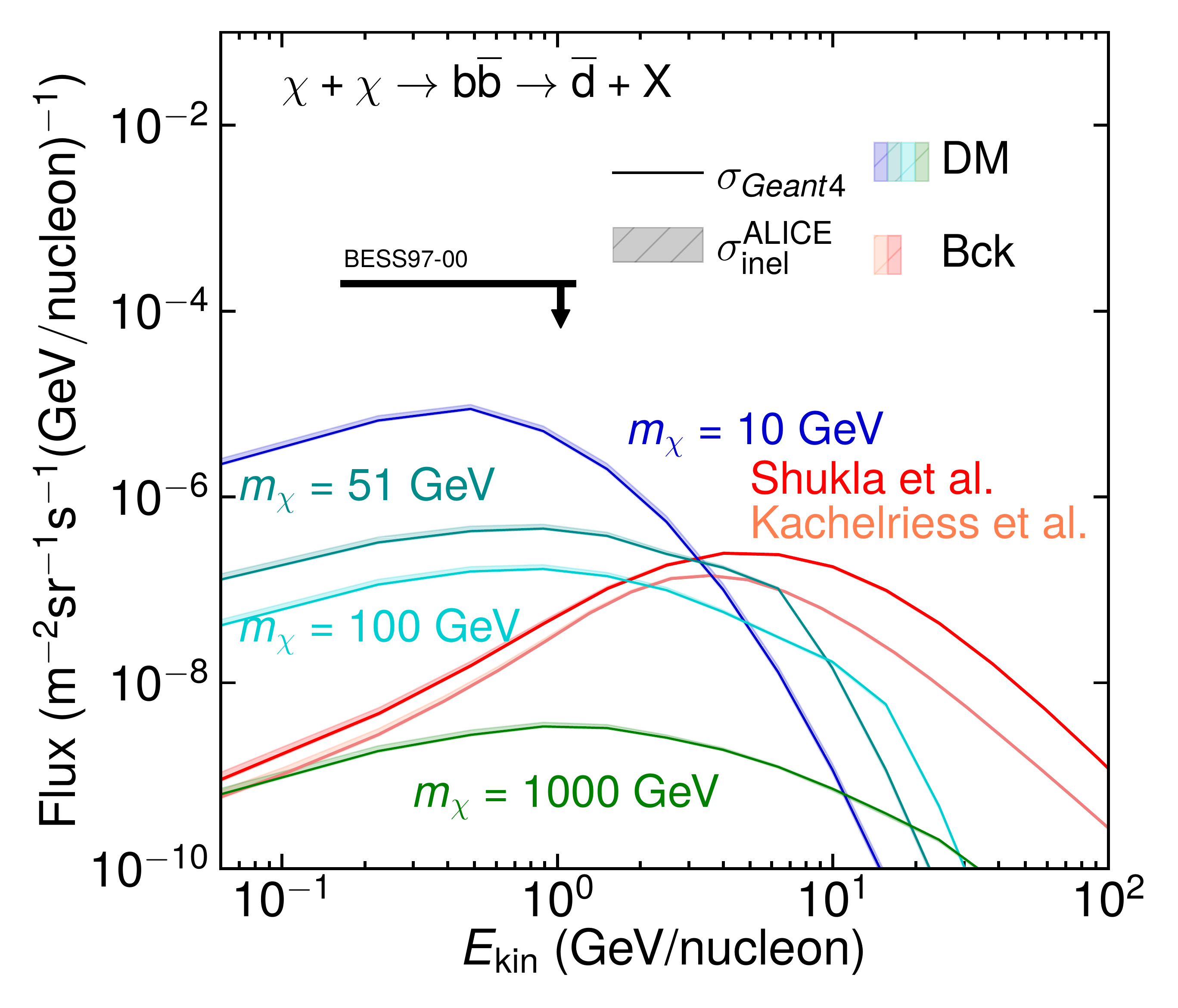}
\includegraphics[width=0.45\textwidth]{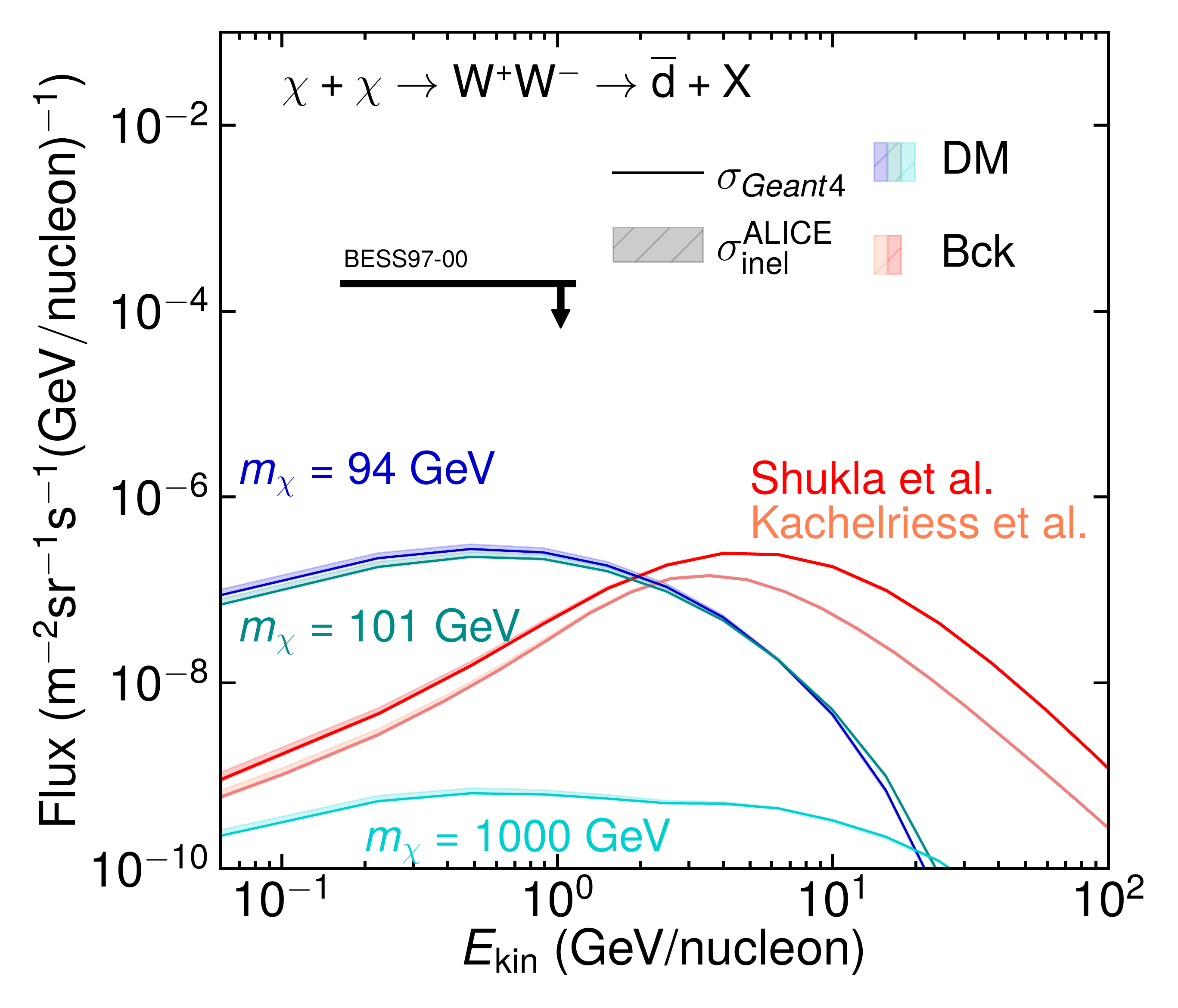}
\includegraphics[width=0.45\textwidth]{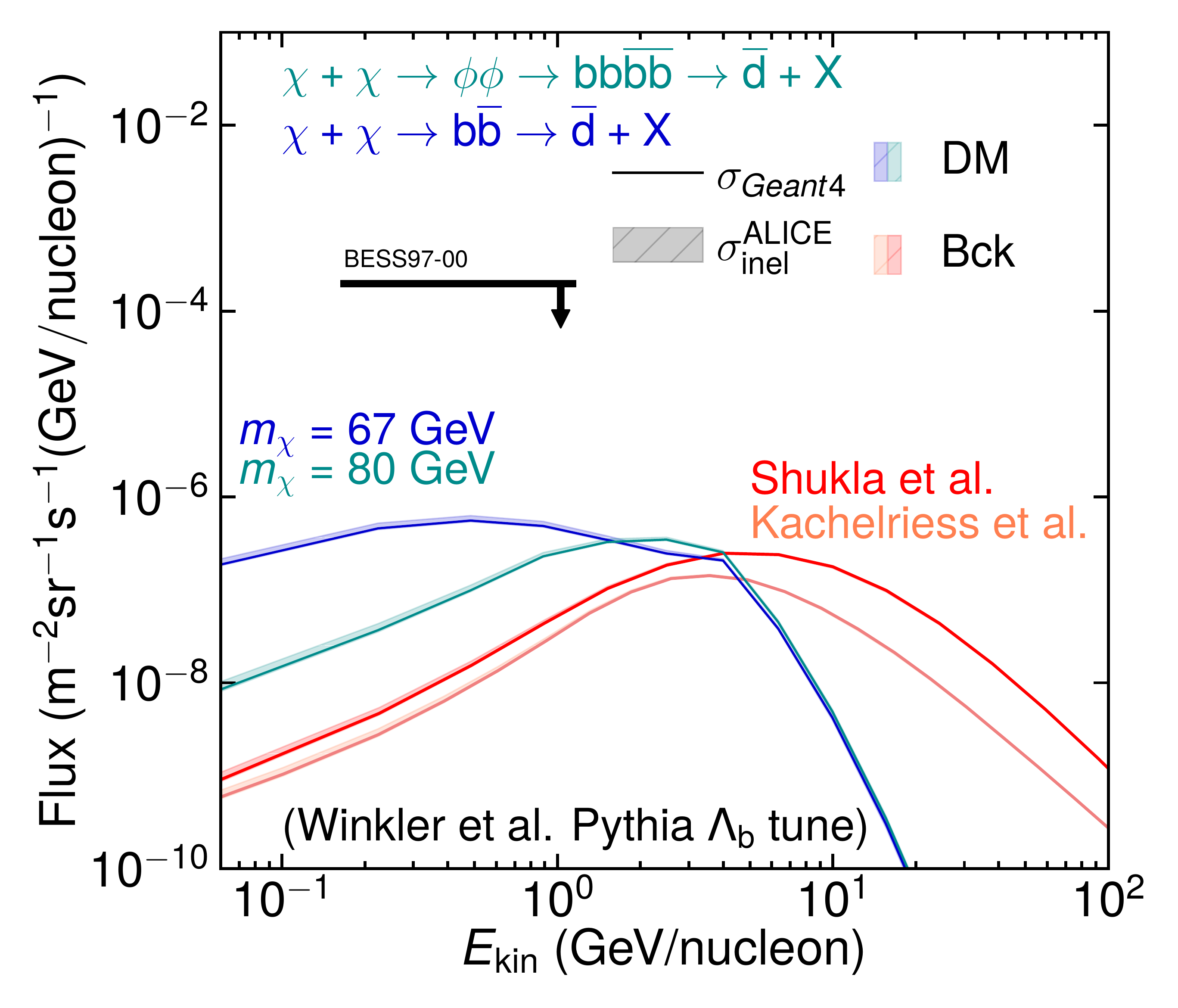}
\includegraphics[width=0.45\textwidth]{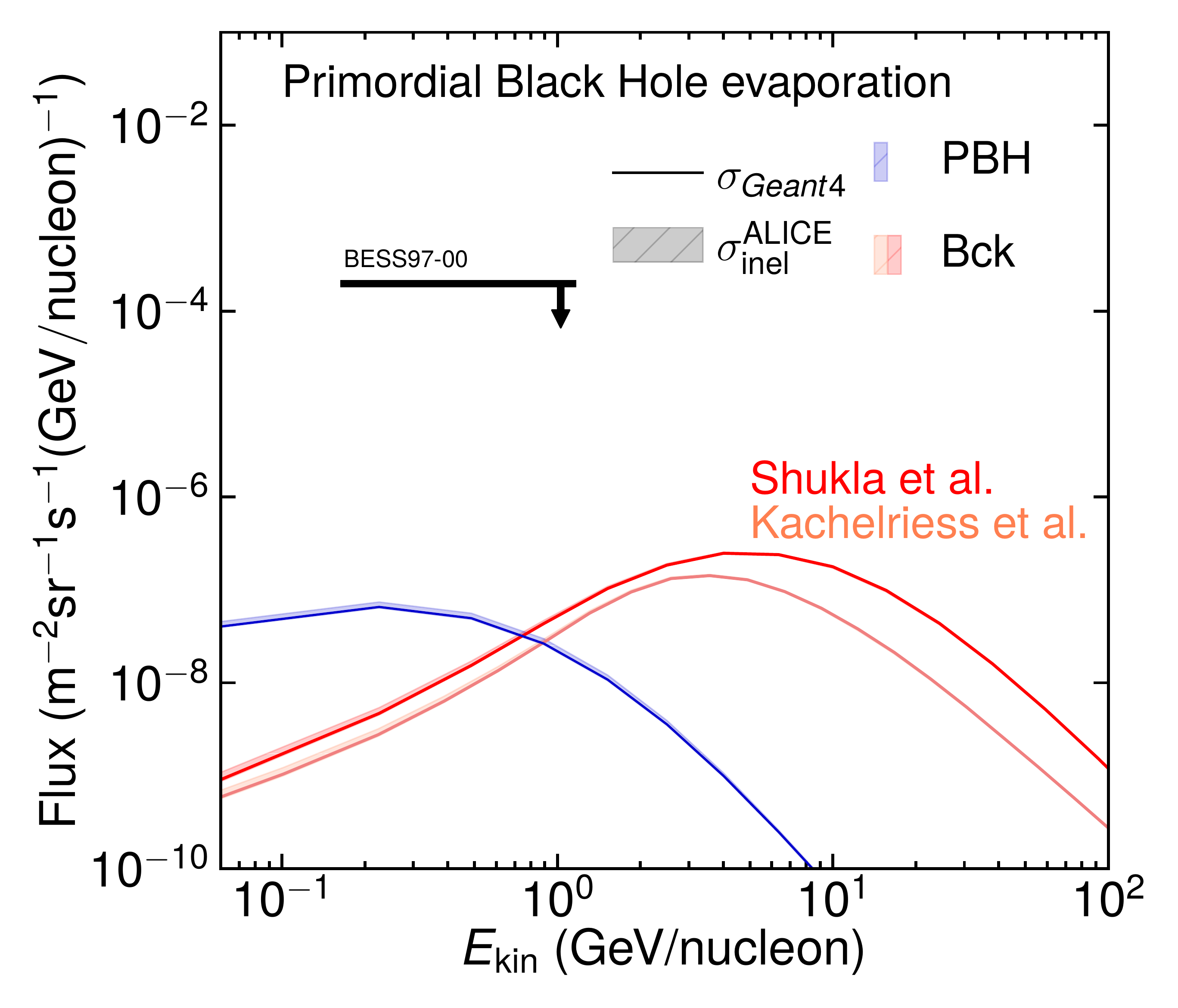}
\caption{Antideuteron fluxes from cosmic-ray collisions with the interstellar medium and for different production scenarios from dark-matter annihilation expected at Earth. Only  uncertainties accounting for the inelastic cross section \sigmainel are shown.} \label{fig:antifeuteronTOA}
\end{figure*}

The predicted antideuteron fluxes at Earth for different sources are shown in Fig.~\ref{fig:antifeuteronTOA}, using fiducial values for source and propagation parameters. The horizontal black line shows the upper limit for the antideuteron flux obtained by the BESS experiment~\cite{Fuke:2005it}.
All panels show the secondary antideuteron flux produced in the collisions of cosmic rays in the interstellar medium. For secondary antideuterons, two coalescence models are considered: Shukla et al.~\cite{shukla_prd} (red) and Kachelrie{\ss} et al.~\cite{Kachelriess:2020uoh} (orange).
%The upper left panel presents results for the antideuterons stemming from dark matter annihilations through the $b\overline{b}$ channel for different dark matter masses ($m_{\chi}=\,10,\,51,\,100,\,1000$ GeV). The dark matter fluxes are represented by the lines in different shades of blue and green.
%The red and orange lines in all panels refer to the secondary antideuterons produced in the collisions of cosmic rays with the interstellar medium. Two models are considered: Shukla et al.~\cite{shukla_prd} (red) and Kachelriess et al.~\cite{Kachelriess:2020uoh} (orange).
The upper two panels of Fig.~\ref{fig:antifeuteronTOA} show antideuterons from dark-matter annihilation into b$\overline{\rm b}$ (left) and W$^+$W$^-$ (right) for several dark-matter masses, using antideuteron spectra determined by~\cite{Ibarra:2012cc}.
%On the upper right panel, the cosmic ray fluxes are shown for the antideuterons stemming from dark matter annihilation via $W^+W^-$ channel, while the 
The lower left panel shows the antideuteron fluxes obtained using production cross sections from Winkler et al.~\cite{Winkler:2020ltd} for the $\bar{\Lambda}_{\rm b}$ decay assumption as explained in section~\ref{sec-DMsources}.
The lower right panel includes the antideuteron flux from primordial black hole evaporation.
The solid lines were obtained using default \geant\ parameterization values for both the dark-matter and cosmic-ray induced fluxes. The shaded bands were obtained with inelastic cross sections estimated using ALICE measurement as described in section~\ref{s-cross}. This effect is discussed in the next section~\ref{sec:inel-uncertainty} and translates into a small flux variation in comparison with the other unknowns entering the calculation. 

Figure~\ref{fig:antifeuteronTOA} leads to the well-known conclusion that low-energy antideuterons ($E_{\rm{kin}}\leq $1\,GeV/nucleon) can exhibit a large signal-to-background ratio between exotic sources of antinuclei and secondary antideuterons. This holds independently of the coalescence model used for the secondary antideuteron flux for both GeV-scale annihilating dark matter and primordial black hole evaporation.

In the following, the different uncertainties entering the calculation of these fluxes are discussed,  taking account of how well they are characterized and their relative importance in predicting the anti\-deuteron flux.

\subsection{Discussion of uncertainties}
The limited knowledge of antideuteron production and propagation hampers the precise prediction of the local antideuteron flux.
Equally important as the fiducial fluxes shown in Fig.~\ref{fig:antifeuteronTOA} is the  accounting for and discussion of the relevant uncertainties:
The uncertainty due to inelastic scattering during propagation is quantified for the first time in section~\ref{sec:inel-uncertainty}. This experimental uncertainty is now several orders of magnitude smaller than the uncertainties due to antideuteron production (see section~\ref{sec:coal-uncertainty}) and diffusion in the Galactic magnetic field (section~\ref{sec:prop-uncertainty}), as well as additional unknowns related to exotic sources of cosmic antideuterons (section~\ref{sec:DM-uncertainty}). All figures in this subsection show the local interstellar fluxes.

%The uncertainty for the production of secondary antideuterons can be gauged by comparing the red and orange fluxes in Fig.~\ref{fig:antifeuteronTOA} and considering that each model beares an intrinsic uncertainty linked to the assumed coalescence parameters. This additional uncertainty is displayed in the right panel of Fig.~\ref{fig:dbar_source}, and it translates linearly to an upper uncertainty on the flux of 27\% and lower uncertainty of 42\%. 

\subsubsection{Loss of antideuteron flux via inelastic interactions\label{sec:inel-uncertainty}}

\begin{figure*}[htbp]
\centering
\includegraphics[width=0.45\textwidth]{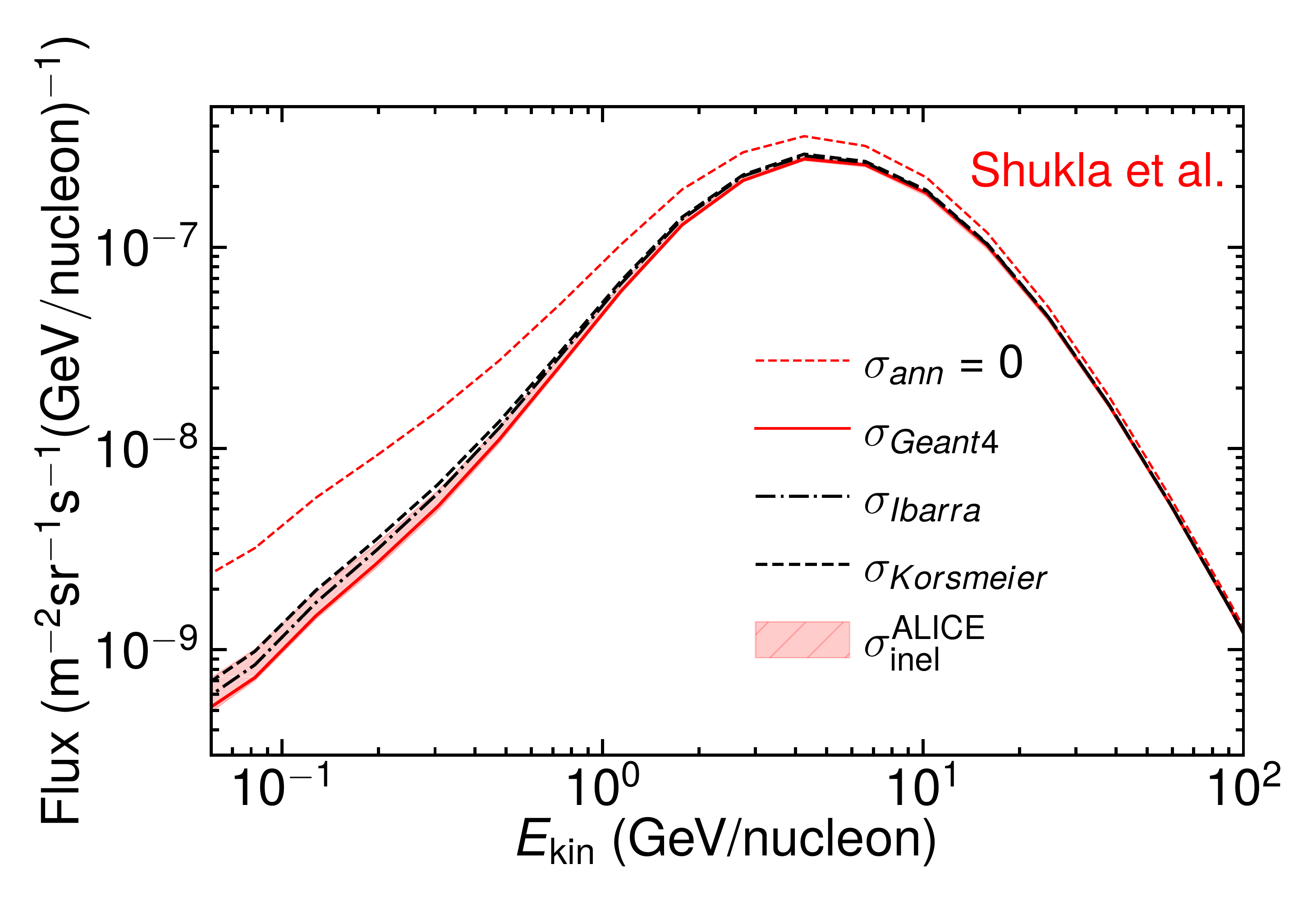}
\includegraphics[width=0.45\textwidth]{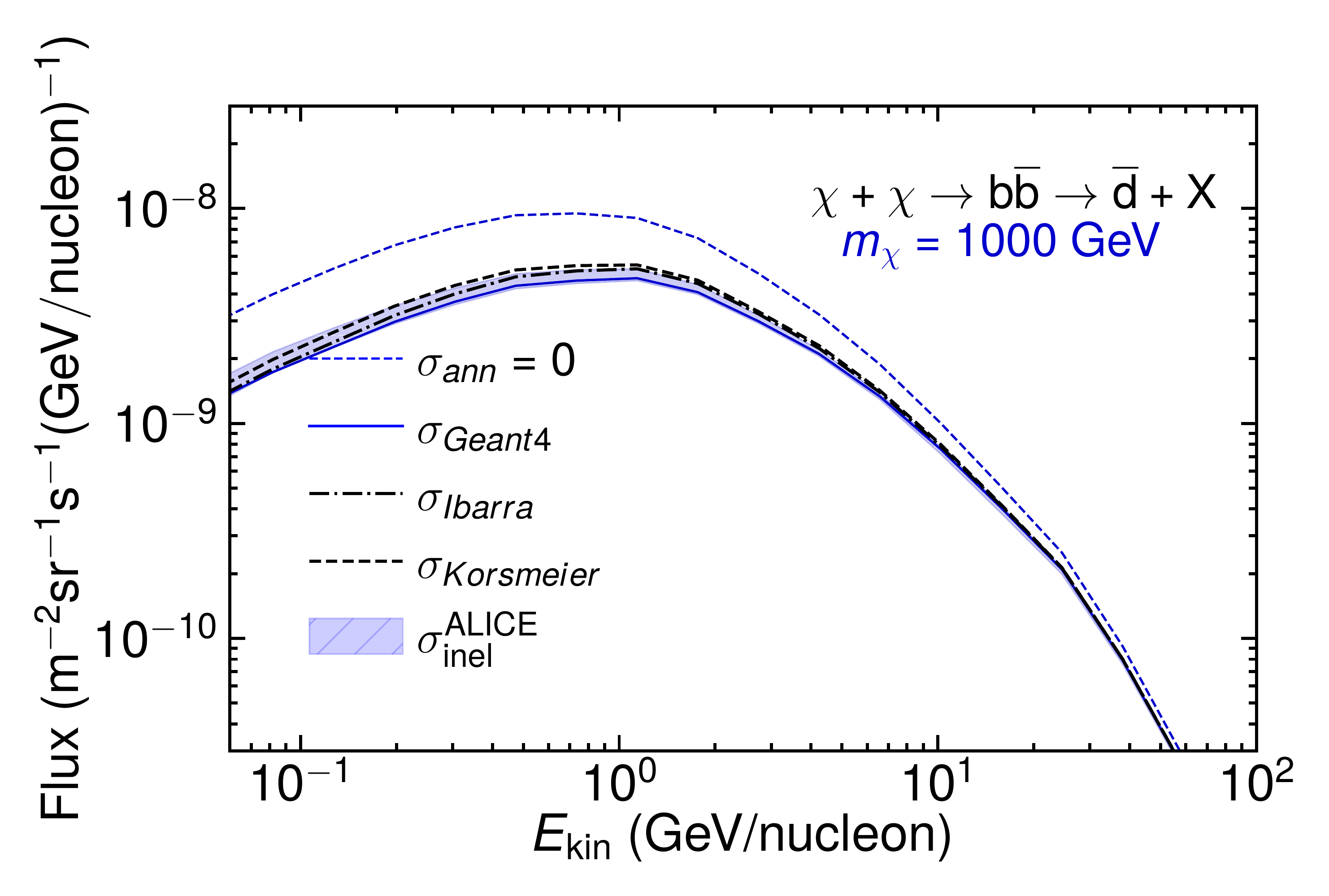}
\includegraphics[width=0.45\textwidth]{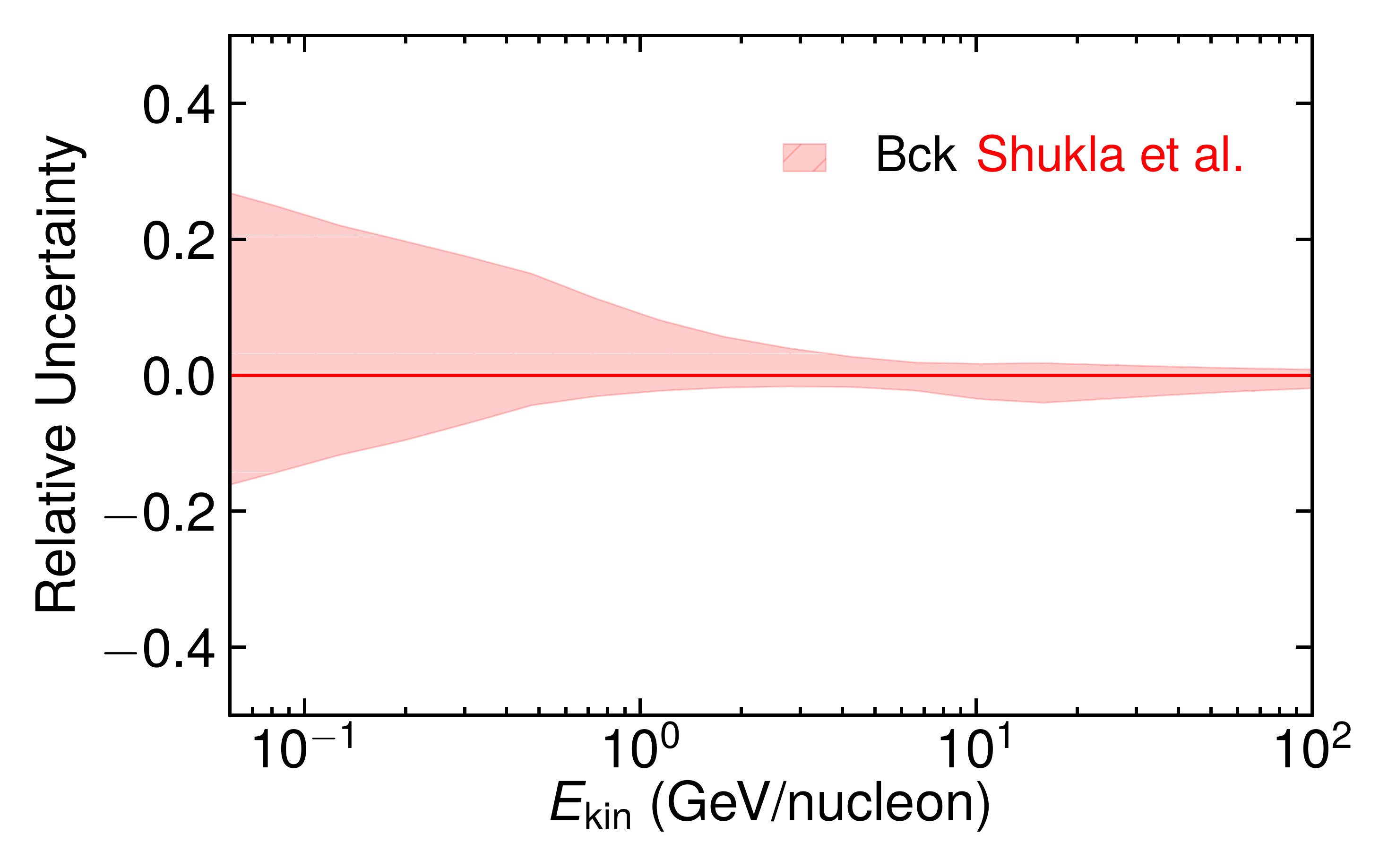}
\includegraphics[width=0.45\textwidth]{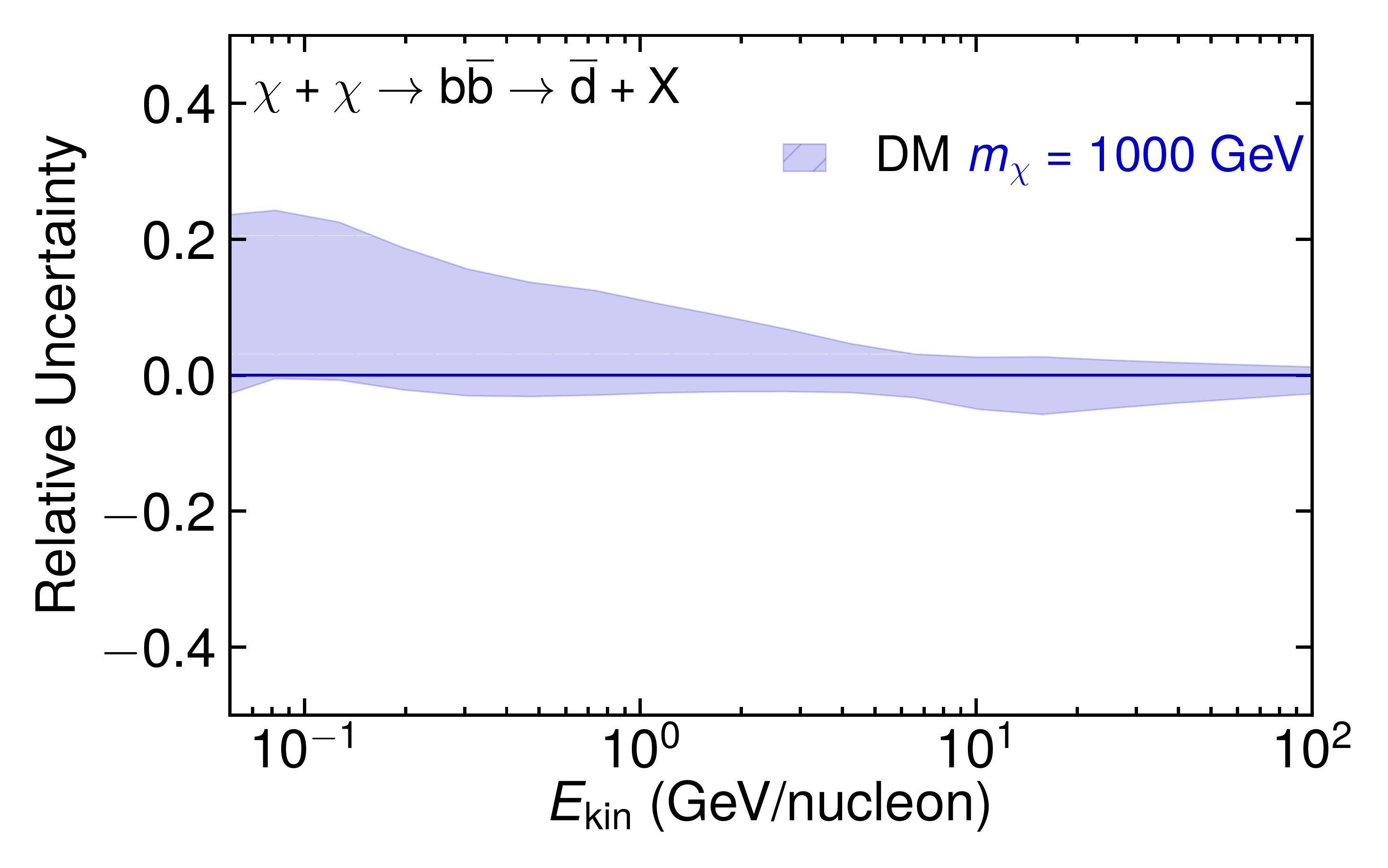}
\caption{Obtained fluxes for exemplary secondary and dark-matter models (top panels) and relative uncertainties obtained using data-driven estimates (bottom panels). The relative uncertainty is shown only from inelastic cross section and before solar modulation.} \label{fig:InelUnc}
\end{figure*}
While the effect of inelastic losses during propagation on the final spectra is modest, it is essential to emphasize that the related uncertainty is now well quantified based on experimental data. 
The result for the inelastic antideuteron scattering cross section presented in section~\ref{s-cross} is not radically different from previous determinations in terms of magnitude, but for the first time allows us properly to quantify the uncertainty in antideuteron flux predictions from inelastic losses. 
The effect of inelastic processes that cause the disappearance of antideuterons on the total flux near Earth for the two different sources is shown in the upper panels of Fig.~\ref{fig:InelUnc}. The left and right upper panels show the expected flux for secondary antideuterons and antideuterons from dark-matter annihilations considering a mass $m_{\chi}=\, 1000$ GeV for different assumptions on the inelastic interactions, respectively. The assumptions are: no inelastic interactions at all (dashed lines) and different parameterizations of inelastic antideuteron cross sections taken from \geant (solid line), from~\cite{Ibarra:2012cc} (dot-dashed lines), from~\cite{Korsmeier:2017xzj} (long-dashed lines) and considering the available experimental measurements~\cite{ALICE:2020zhb,Denisov:1971im,Binon:1970yu} (coloured bands). Details of the different parameterizations are described in section~\ref{s-cross}. \\
One can see that fluxes calculated using the measured inelastic cross sections agree with previous parameterizations obtained from a scaling of the measured antiproton inelastic cross sections within the estimated uncertainties. Furthermore, the effect of the absorption on the calculated fluxes is most prominent at lower kinetic energies where the inelastic cross section is maximal (see Fig.~\ref{fig:sigmainel_results}).
The uncertainty on the inelastic cross section does not translate linearly to the uncertainty on fluxes, as can be seen in the lower panels of Fig.~\ref{fig:InelUnc}, where this relative uncertainty is shown as a function of the antideuteron kinetic energy for the secondary (left lower panel) and dark-matter sources (right lower panel). 
%\jc{alternative to following 2 paragraphs:}
In general, the effect of inelastic scattering becomes stronger for longer propagation times. It is small at large energies, where cosmic-ray propagation is mainly escape-dominated, and larger at small energies. This holds in particular for secondary antideuterons, where energy losses are responsible for the reduction in flux below the production threshold.

\subsubsection{Production uncertainty}
\label{sec:coal-uncertainty}
The dominant uncertainty in predicting the antideuteron flux is related to our imperfect knowledge of antideuteron production in high-energy processes, see section~\ref{sec:coalescence}.
Within a given coalescence model, the model parameters (the coalescence momentum $p_0$ in~\cite{Gomez-Coral:2018yuk,shukla_prd} and the size of emission region $\sigma$ in~\cite{Kachelriess:2020uoh}) are determined from fitting to antideuteron production data, and their plausible ranges can be determined.
For secondary antideuterons, this is shown as red and orange bands in Fig.~\ref{fig:antifeuteronTOA}. The effect of adopting a smaller or larger coalescence momentum on the antideuteron yield is roughly independent of energy. This can be understood in terms of the separation of scales, where changing the coalescence condition at small energies does not significantly impact the overall energy distribution of the antinucleon pairs produced at higher energy.
The quantified coalescence uncertainty hence amounts to a simple rescaling of the fluxes displayed in Figure~\ref{fig:antifeuteronTOA}.
For secondary antideuterons, this can be determined from data on antideuteron production in collisions of nuclei, resulting in $\!^{+27}_{-42}\%$ uncertainty on the flux for Shukla et al.~\cite{shukla_prd}, and $\pm 20\%$ for Kachelrie{\ss} et al.~\cite{Kachelriess:2020uoh}. As there are no experimental data on the hypothetical processes of dark-matter annihilation or PBH evaporation, the situation is less clear in this case. The $p_0$ ranges required to reproduce antideuteron production at different Standard Model processes thought to most closely resemble dark-matter annihilation, or PBH evaporation (i.e., hard processes producing $\bar{\rm q}$q pairs in isolation) do not agree with each other. Taking their envelope, one obtains a plausible range of $\!^{+63}_{-70}\%$ for the dark-matter annihilation and PBH evaporation fluxes following~\cite{Ibarra:2012cc}, which can, however, no longer be interpreted as a meaningful uncertainty band.

More important than the uncertainty resulting from the determination of coalescence parameters within any given model may be the systematic uncertainty from imperfect modeling. For the secondary flux, this is evident in Fig.~\ref{fig:dbar_source}, where the two different secondary predictions  often do not overlap within their uncertainty bands, with the estimated upper limit from the Shukla et al.\ model being up to 7 times larger than the lower limit obtained employing the Kachelrie{\ss} et al.\ model.
The same is true for exotic sources of antideuterons, where in particular, there has been discussion regarding the possibility of increased antinuclei production in $\bar{\Lambda}_{\rm b}$-decay~\cite{Winkler:2020ltd,Kachelriess:2021vrh,Winkler:2021cmt}. Therefore, the microscopic modeling of antideuteron production needs to be improved to reduce the current uncertainties, which requires measuring antideuteron production with accelerator experiments at different energies in different production channels.
This scheme should be tuned for collisions at intermediate energies that match the energy scale of the processes induced by cosmic rays. The studies of the antinuclei formation arising from charm-hadron decays could help in better constraining possible dark-matter decays.
\subsubsection{Propagation parameters}
\label{sec:prop-uncertainty}
Propagation models are constrained by measurements of several primary and secondary cosmic-ray species.
The parameters obtained by Boschini et al.\ were used as defaults in this work. We also compute fluxes using parameters by Cuoco et al.~\cite{PhysRevLett.118.191102} to illustrate the uncertainty related to propagation. Both sets of propagation parameters are summarized in Table~\ref{tab:tablePar}. Boschini et al.\ used Voyager 1, AMS-02, HEAO-3-C2, and ACE-CRIS experimental data to fit the propagation parameters~\cite{2020ApJS..250...27B} while Cuoco et al.\ employed Voyager~1, AMS-02, and CREAM data~\cite{PhysRevLett.118.191102}.

\begin{figure}[htbp]
\centering
\includegraphics[width=0.45\textwidth]{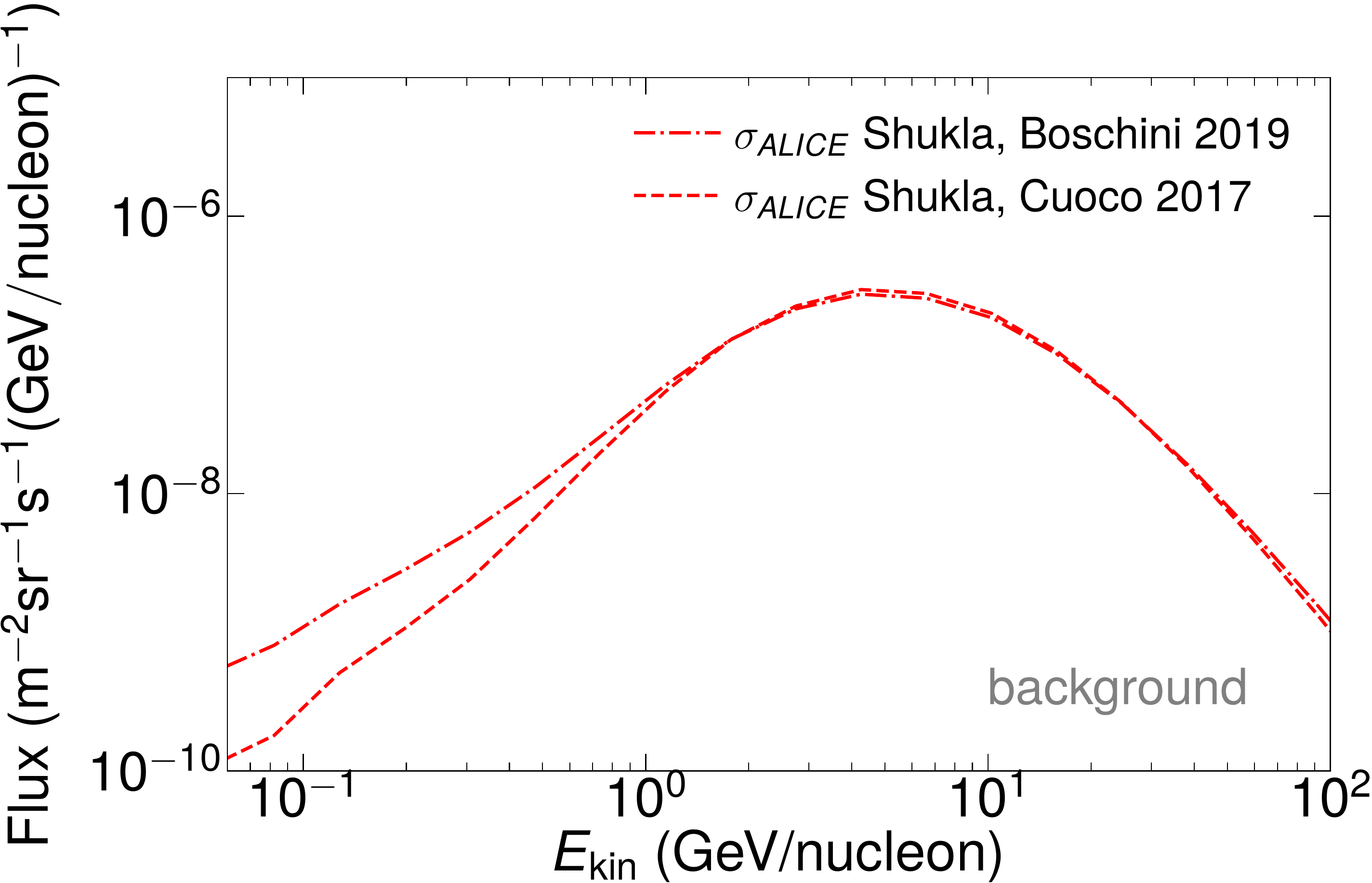}
\includegraphics[width=0.45\textwidth]{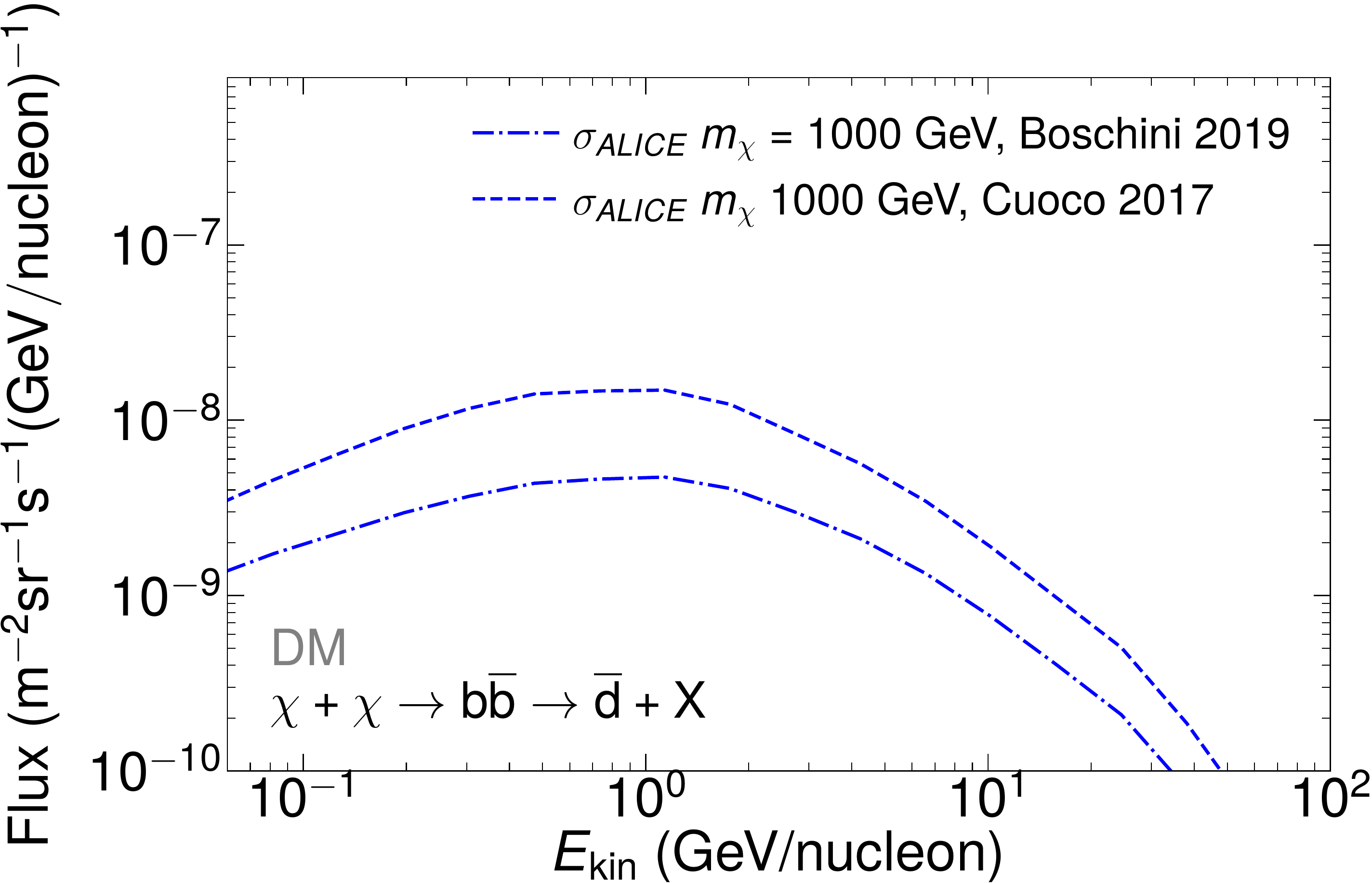}
\caption{The antideuteron flux obtained using two different sets of propagation parameters.} \label{fig:dependenceOnPropagation}
\end{figure}
Fig.~\ref{fig:dependenceOnPropagation} compares antideuteron fluxes obtained with these two propagation benchmarks.
While propagation parameters in the two works are rather different, the computed fluxes of secondary antideuterons at higher energies ($E_{\rm{kin}} \geq $1 GeV/nucleon) are in good agreement (see upper panel of Fig.~\ref{fig:dependenceOnPropagation}). This is expected since both benchmarks were built to reproduce the available AMS-02 data, which constrains this energy regime very well. The difference at low energies can be attributed to the stronger convection effects assumed by Cuoco et al.

In the case of the antideuteron flux arising from dark-matter annihilation (see lower panel of Fig.~\ref{fig:dependenceOnPropagation}), the flux obtained using the propagation parameters from Cuoco et al. is about 2--3 times larger at all energies than that using the Boschini et al. parameters. This can be explained by the different halo half-width values in the parameterizations ($z_{\rm{h}}$ parameter in Table~\ref{tab:tablePar}). 
It is well known that there is a degeneracy between the height of the diffusive halo and the diffusion coefficient when predicting secondary cosmic-ray fluxes.
However the dark-matter halo extends well beyond the diffusive halo, and the larger diffusive halo results in a larger number of diffusively confined dark-matter annihilation products.
This enhancement of the dark-matter induced antideuteron flux with larger halo size does not depend strongly on energy and is the same for antiprotons and antideuterons.
Hence, the associated uncertainty largely cancels when predicting antideuteron fluxes based on antiproton signals or limits.

\subsubsection{Dark matter specific uncertainties}
\label{sec:DM-uncertainty}
The possible antideuteron flux arising from dark-matter annihilation is not well constrained since knowledge of the hypothetical dark-matter particles is very limited. For example, the dark-matter mass $m_\mathrm{DM}$ is not known.
For the velocity-averaged annihilation cross section $\langle\sigma v\rangle$, there is the well-motivated thermal WIMP benchmark value $\langle\sigma v\rangle_\mathrm{thermal}$, from which it can differ by orders of magnitude. This can be both due to effects particular to any given WIMP model (e.g., velocity suppressed annihilation cross section, Sommerfeld enhancement) or astrophysical boost factors due to dark-matter clumping (eg.~\cite{Stref:2016uzb}).
Still, results for different dark-matter masses and annihilation channels together give a qualitative picture of what kind of dark-matter-induced antideuteron fluxes can be expected (cf.\ Fig~\ref{fig:antifeuteronTOA}). 
\begin{figure}[htbp]
\centering
\includegraphics[width=0.45\textwidth]{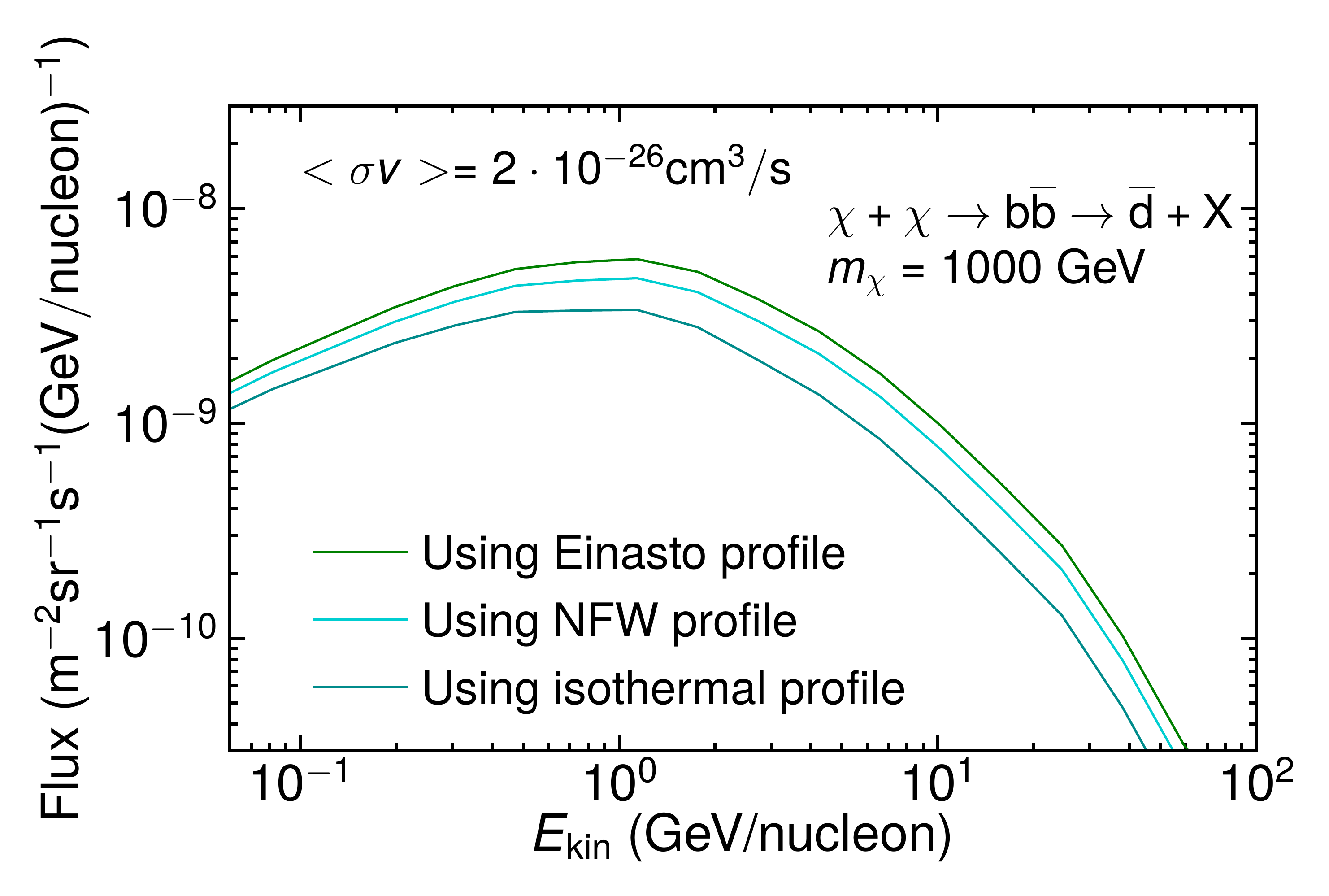}
\caption{Antideuteron fluxes from dark-matter annihilations for different dark-matter density profiles. } \label{fig:transparency_vs_DM_profiles_log}
\end{figure}
The uncertainty arising from the choice of the dark-matter density profile is shown in Fig.~\ref{fig:transparency_vs_DM_profiles_log}.
%where the different lines represent antideuteron cosmic-ray fluxes obtained using different dark-matter density profiles. 
The isothermal profile shown in Fig.~\ref{fig:DM_profiles_log} assumes a smaller density of dark matter in the inner Galaxy than the NFW profile and thus results in 14\%-62\% smaller antideuteron fluxes, as seen in Fig.~\ref{fig:transparency_vs_DM_profiles_log}. On the other hand, as the Einasto profile assumes a higher central density of  dark matter than NFW, it results in 12\%-44\% higher fluxes.
As can be seen in Fig.~\ref{fig:transparency_vs_DM_profiles_log}, the resulting difference is largely a normalization effect.
Such normalisation uncertainties shared by antiprotons and antideuterons from DM cancel exactly when predicting antideuteron fluxes based on antiproton signals or limits.

\section{Summary}
In summary, updated antideuteron fluxes at Earth for several dark-matter masses and two models for the secondary production due to cosmic rays have been presented, taking into account for the first time the measurement of the antideuteron inelastic cross section. The error associated with this measurement is propagated to the antideuteron flux and we show that  because of the new experimental measurement this component of the modelling is very well constrained and has negligible uncertainty with respect to the other uncertainty sources. %{\color{magenta}AI: too negative? One could say instead that this source of uncertainty has been greatly reduced, but that the expected fluxes still suffer from large uncertainties from the modelling of the dbar formation or from progagation. (we confirm that the inel cs is not the main unceratainty)} 

The results have been evaluated following a consistent scheme, in which all fluxes are obtained using the same propagation model and the same inelastic cross section. Thus it is possible to compare the two considered models for the secondary sources quantitatively.  
A detailed discussion of the uncertainties related to the predicted fluxes show that the major contributions are currently associated to propagation parameters, to the microscopic modelling of antinuclei formations in both cosmic-rays collisions and dark-matter decays and to the still unconstrained dark-matter modelling.

The current work represents a state-of-the-art method for antideuteron predictions that could be further advanced only by improving  the experimental studies of light nuclei formation and their interpretation, to be applied to both dark-matter and secondary sources, and on the extension of the kinematic range for the inelastic cross section measurements. The microscopic behaviour of  nuclei production should especially  be understood in the energy range most relevant for the astrophysical processes.
%One study that could be performed to improve on the current study is the "full chain" we discussed, where one fits to AMS antiproton data and propagation at the same time, in order to get the "maximum possible" antideuteron flux. I therefore added the work "method" to allow for this possibility. 

The potential of antideuterons as a messenger is not changed by this work.
While the effect of inelastic losses during propagation on the final spectra is modest, it should be emphasised that the related uncertainty is now well quantified based on experimental data.  
Further studies at accelerator will help in improving the predictions and interpreting future antideuteron signals by satellite or balloon experiments.

\section{Acknowledgement}
We are grateful to M.\ Kachelrieß and J.\ Tjemsland for providing us with their antideuteron production cross sections. We thank as well M.\ Winkler for providing us the antideuteron spectra from dark matter annihilation.

P.~von~Doetinchem, D.~Gomez, and A.~Shukla received support from the National Science Foundation under award PHY-2013228. This research was done using resources provided by the Open Science Grid~\cite{osg07, osg09}, which is supported by the National Science Foundation Grant No. 1148698, and the U.S. Department of Energy's Office of Science. The technical support and advanced computing
resources from the University of Hawaii Information Technology Services -- Cyberinfrastructure are gratefully acknowledged.

This work is supported by the Deutsche Forschungsgemeinschaft (DFG, German Research Foundation) through Grant SFB 1258 ``Neutrinos and Dark Matter in Astro- and Particle Physics'' and through Excellence Cluster ORIGINS under Germany's Excellence Strategy - EXC-2094-390783311.

\bibliography{main}% Produces the bibliography via BibTeX.

\end{document}